\documentstyle[11pt,aaspp4]{article}
\def\etal{{et al.\hskip 3pt}}
\def\ie{{i.e.\hskip 3pt}}
\def\eg{{e.g.\hskip 3pt}}

\newcommand{\lya}{Ly$\alpha$\ }
\newcommand{\dndz}{$dN/dz$}
\newcommand{\dndw}{$dN/dW_r$}
\newcommand{\nh}{N_{\rm HI}}
\newcommand{\nhl}{N_{\rm HI,lim}}
\newcommand{\kms}{\;{\rm km}\,{\rm s}^{-1}}

\newcommand{\hkpc}{h^{-1}\;{\rm kpc}}
\newcommand{\hmpc}{h^{-1}\;{\rm Mpc}}
\newcommand\cdunits{{\rm cm}^{-2}}
\newcommand\ggh{\Gamma_{\rm HI}}

\slugcomment{Submitted to ApJ}

\lefthead{Dav\'e et al.}
\righthead{Evolution of \dndz}
\singlespace

\begin{document}

\title{The Low Redshift Lyman Alpha Forest in Cold Dark Matter Cosmologies}

\author{Romeel Dav\'e and Lars Hernquist}
\affil{Astronomy Department, University of California, Santa Cruz, CA 95064}

\author{Neal Katz}
\affil{Astronomy Department, University of Massachusetts, Amherst, MA 01003}

\and

\author{David H. Weinberg}
\affil{Astronomy Department, Ohio State University, Columbus, OH 43210}

\begin{abstract}

We study the physical origin of the low-redshift \lya forest
in hydrodynamic simulations of four cosmological models,
all variants of the cold dark matter scenario.  Our most
important conclusions are insensitive to the cosmological model, but they
depend on our assumption that the UV background declines at low redshift
in concert with the declining population of quasar sources.
We find that the expansion of the universe drives rapid evolution of \dndz\ 
(the number of absorbers per unit redshift above a specified equivalent 
width threshold) at $z \ga 1.7$, but that at lower redshift
the fading of the UV background counters the influence of expansion, leading
to slow evolution of \dndz.  The draining of gas from low density regions 
into collapsed structures has a mild but not negligible effect on the
evolution of \dndz, especially for high equivalent width thresholds.
At every redshift, weaker lines come primarily from
moderate fluctuations of the diffuse, unshocked intergalactic
medium (IGM), and stronger lines originate in shocked or radiatively cooled
gas of higher overdensity.  However, the neutral hydrogen column density
associated with structures of fixed overdensity drops as the universe expands,
so an absorber at $z=0$ is dynamically analogous to an absorber that has
column density 10 to 50 times higher at $z=2-3$.
In particular, the mildly overdense IGM fluctuations that dominate the \lya
forest opacity at $z>2$ produce optically thin lines 
at $z<1$, while the marginally saturated ($\nh \sim 10^{14.5}\;\cdunits$)
lines at $z<1$ typically arise in gas that is overdense by a factor of $20-100$.
We find no clear distinction between lines arising 
in ``galaxy halos'' and lines arising in larger scale structures; however,
galaxies tend to lie near the dense regions of the IGM that are responsible
for strong \lya lines.
The simulations provide a unified physical picture that accounts for the 
most distinctive observed properties of the low-redshift \lya forest:
(1) a sharp transition in the evolution of \dndz\ at $z \sim 1.7$, (2) stronger
evolution for absorbers of higher equivalent width, (3) a correlation of
increasing \lya equivalent width with decreasing galaxy impact parameter that
extends to $r_p \sim 500\hkpc$, and (4) a tendency for stronger lines 
to arise in close proximity to galaxies while weaker lines trace more
diffuse large scale structure.

\end{abstract}


\section{Introduction}\label{sec: intro}

\lya absorption by neutral hydrogen
along the line of sight to distant quasars provides a sensitive
probe of the cosmic distribution of diffuse gas over a wide range of redshifts 
(\cite{bah65}; \cite{gun65}; \cite{sche65}; \cite{lyn71}; \cite{sar80}).
In recent years, the Keck telescope's HIRES spectrograph (\cite{vog94}) 
has revolutionized the study of the \lya ``forest"
at redshifts $z \sim 2-4.5$, yielding
high-resolution spectra of spectacular precision (see, e.g.,
\cite{hu95}; \cite{lu96}; Kim \etal 1997, hereafter \cite{kim97}; 
\cite{kir97}).
At the same time, the UV spectroscopic capabilities of
the Hubble Space Telescope (HST) have made it possible to study the
\lya forest from $z=2$ down to $z=0$.  
In particular, the HST Quasar Absorption Line Key Project team has used the 
Faint Object Spectrograph (FOS) to conduct a comprehensive census
of low redshift \lya lines, especially those
with rest-frame equivalent width $W_r \geq 0.24$\AA,
examining statistical quantities such as the distribution of
equivalent widths, \dndw, and the evolution of the number
of lines per unit redshift above an equivalent width threshold, \dndz\
(\cite{bah93}, 1996; \cite{jan98b};
Weymann \etal 1998, hereafter \cite{wey98}).  Other groups
have used the Goddard High Resolution Spectrograph (GHRS) 
to study weaker lines toward a smaller number of quasars
(e.g., \cite{mor91}; \cite{sto95}; Shull, Stocke \& Penton 1996, 
hereafter \cite{shu96}; Tripp, Lu, \& Savage 1998, hereafter
\cite{tri98}).
Future observations with the Space Telescope Imaging Spectrograph (STIS)
and the Cosmic Origins Spectrograph (COS) should yield more extensive
samples of weak low-redshift lines.

Concurrently with these observational developments, cosmological simulations
that incorporate gravity, gas dynamics, radiative cooling, and 
photoionization have shown that theoretical models based on inflation
and cold dark matter (CDM) can reproduce many of the 
observed properties of the high-redshift ($z>2$) \lya forest
(e.g., \cite{cen94}; \cite{zha95}; \cite{her96}; \cite{mir96}; 
\cite{wad97}; \cite{bry98}; \cite{the98}b).  
These theoretical models were originally
introduced to account for the observed properties of galaxies, large scale 
structure, and cosmic microwave background anisotropies, so their
success in explaining an entirely different class of observational phenomena
provides important support for at least the broad features of the CDM scenario.
The simulations lead to a physical picture in which most of the
optically thin or marginally saturated lines at high redshift
(those with neutral hydrogen
column densities $\nh \la 10^{14.5}\cdunits$) arise in structures that
are only a few times the cosmic mean density and are therefore far
from dynamical or thermal equilibrium.  The physics of the diffuse
intergalactic medium (IGM) that produces this low column density absorption
is relatively simple, so this model of the high-redshift \lya forest
can be described with surprising accuracy by straightforward analytic
or numerical approximations (e.g., \cite{bi93}; \cite{bi95}; 
\cite{muc96}; \cite{bi97}; \cite{hui97}; \cite{cro98}; \cite{gne98}; 
Weinberg, Katz, \& Hernquist 1998b).

In this paper we use smoothed particle hydrodynamics (SPH) simulations
to extend these numerical studies to $z=0$, employing a newly developed
parallel version (\cite{dav97a}) of our simulation code 
TreeSPH (\cite{her89}; Katz, Weinberg, \& Hernquist 1996, 
hereafter \cite{kat96}).
We consider four variants of the CDM scenario 
(\cite{peebles82}; \cite{blu84}), two with $\Omega=1$ and two with
$\Omega=0.4$.  However, in this paper we focus on 
understanding the physics of the low-$z$ forest rather than distinguishing
between cosmological models.  The \lya forest is a promising new arena
for cosmological tests, but these require
a careful match between theoretical and observational
analyses that is beyond the scope of the present study.
We plan to pursue such comparisons at both low and high redshift in 
future work.  At various points in the paper, we will compare our
results to those of other numerical investigations of the low-redshift forest:
Riediger, Petitjean, \& M\"ucket (1998),
who use a modified N-body method, and Theuns, Leonard, \& Efstathiou (1998a),
who use SPH simulations.

A surprise in the very first spectra of the low-redshift \lya forest
was the discovery of many more lines than expected from an extrapolation
of high-redshift data (\cite{bah91}; \cite{mor91}).  The implied change
in the rate of evolution of the forest has now been precisely
quantified by the Key Project team (\cite{wey98}).  Adopting
Sargent et al.'s (1980) parameterization of the evolution,
\begin{equation}\label{eqn: gamma}
{dN\over dz}= \left({dN\over dz} \right)_0 (1+z)^\gamma ~,
\end{equation}
\cite{wey98} find $\gamma = 0.26 \pm 0.22$ for a sample of nearly
500 lines with $W_r \geq 0.24$\AA\ in the spectra of 62 quasars with $z \la 1.5$.
Ground-based studies imply much steeper evolution at $z>2$
(e.g., \cite{mur86}; \cite{lu91}; \cite{bec94}),
with the most recent results coming from the
Keck/HIRES observations of \cite{kim97}, which yield $\gamma= 2.78\pm 0.71$.
Combining the HST and ground-based
results implies a sharp break in the evolution of \dndz\ 
at $z \sim 1.7$, which is further supported by
\cite{wey98}'s analysis of the quasar UM 18 at $z=1.89$.
The break in evolution has led to the suggestion that there are two
distinct populations of \lya absorbers, one (perhaps associated with
filaments and sheets) evolving rapidly and being dominant at high $z$,
and the other (perhaps associated with galaxy halos) evolving slowly
and being dominant at low $z$ (e.g., \cite{bah93}, 1996,
who also discuss other possible causes for the break in \dndz\ evolution).

The idea of an association between \lya absorbers and gaseous
halos of galaxies has a long history (e.g., Bahcall \& Spitzer 1969).
One important practical virtue of low-$z$ forest studies is that follow-up
observations can examine the relation between \lya absorbers and the
neighboring galaxies and large scale structure.  
For example, Lanzetta \etal (1995, hereafter \cite{lan95}) and 
Chen \etal (1998, hereafter \cite{che98}) have conducted deep
imaging and spectroscopic surveys of galaxies in Key Project fields
and find that galaxies with impact
parameters $r_p < 160\hkpc$ almost always have an associated 
absorption line with $W_r \ga 0.3$\AA, but that galaxies with
$r_p > 160\hkpc$ almost never do ($r_p$ is the projected distance
from the galaxy to the quasar line of sight). 
They also find a trend of
increasing \lya equivalent width with decreasing galaxy impact
parameter, a correlation that has been extended to larger $r_p$ and
smaller $W_r$ by \cite{tri98} (who add a cautionary discussion of
selection biases that could artificially amplify the apparent trend).
On this basis, \cite{lan95} and \cite{che98} argue that most, and perhaps
all \lya absorbers with $W_r > 0.3$\AA\ arise in extended gaseous envelopes
surrounding galaxies, with typical radii $\sim 160\hkpc$
(weakly dependent on galaxy luminosity).
However, numerous studies show
that there are some \lya absorbers with no luminous galaxy nearby
(e.g., \cite{mor93}; \cite{bowen96}; \cite{leb96}; \cite{vgo96};
\cite{leb98}; \cite{tri98})
even with impressively deep imaging (\cite{rau96}).
Comparison to wide-angle galaxy redshift surveys shows
that some weak \lya absorbers reside in large scale voids of
the galaxy distribution (\cite{sto95}; \cite{shu96}).
Statistical analyses show that the low-$z$ \lya absorbers are not
randomly distributed with respect to galaxies, but they also show
that the cross-correlation between absorbers and galaxies is
not as strong as the correlation of galaxies with themselves
(\cite{mor93}; \cite{lan98}; \cite{tri98}) and that
the galaxy density around absorbers is lower than the galaxy density
around galaxies (\cite{gro98}).
These analyses have led many of the above authors (and others as well) 
to suggest that low-$z$ \lya forest absorption is produced not by 
individual galaxies but by larger scale gaseous structures that the galaxies 
themselves reside in, with some tendency for weak absorbers to avoid high 
density regions and favor low density regions.

Given this observational background, we began our theoretical investigation
of the low-$z$ \lya forest with several interlocking questions in mind.
First, can the simulations account for the observed break in \dndz\ 
evolution at $z \sim 1.7$, and if so, what is the mechanism causing
the break?  Second, are the physical structures that produce the \lya
forest the same at high and low redshift, or is there a transition
from one population of absorbers to another?  Third, if the low redshift
picture is similar to the one at high redshifts, how do the simulations
account for the apparent correlations between low-$z$ absorbers and
galaxies, and can they reconcile the somewhat contradictory evidence
on absorber-galaxy associations
supplied by different observational studies?
We will argue below that the simulations provide a good account of
the observed properties of the low-$z$ forest, that the answer to 
the evolution question is fairly simple, and that the answers
to the subsequent questions are clear but multi-faceted.

We describe our simulations and our method of creating
and analyzing artificial spectra in \S\ref{sec: sims}.  In
\S\ref{sec: dndz} we compare \dndz\ and \dndw\ 
computed from our simulated spectra to the
most recent analysis of Key Project data at low redshift (\cite{wey98}) and
to an analysis of Keck HIRES spectra at high redshift (\cite{kim97}).  In
\S\ref{sec: physprop} we investigate the physical properties of the
gas giving rise to \lya absorption.  In \S\ref{sec: gals} we identify
sites of galaxy formation in our simulations, and quantify
relationships between \lya absorbers and galaxies.
Finally, in \S\ref{sec: concl} we summarize our conclusions on
the nature of the \lya forest at low redshifts.  

\section{Simulations and Artificial Spectra}\label{sec: sims}

We employ hydrodynamic simulations of four currently
favored variants of the CDM scenario:
a Lambda-dominated CDM model (LCDM) with $\Omega_\Lambda=0.6$
(\cite{lid96b}), a tilted CDM model with a slope of $n=0.8$ on large
scales (TCDM; \cite{whi96}), an open CDM model (OCDM) with $\Omega=0.5$
(\cite{lid96a}), and a Cold~+ Hot Dark Matter model (CHDM) with two
neutrino species totaling $\Omega_\nu=0.2$ (\cite{kly97}).  The
properties of the models are listed in Table~\ref{table: models}; each
is roughly consistent with COBE normalization, cluster normalization,
and the baryon density $\Omega_b\approx 0.02 h^{-2}$, as determined
from recent measurements of the primordial deuterium abundance (\cite{bur98a},b)
and as favored by the estimates of the mean opacity of the \lya forest
at high redshifts by Rauch et al.\ (1997b).  
The CHDM model has a slightly lower
$\Omega_b h^2$ than the others.

The simulations were performed using Parallel TreeSPH (Dav\'e et al.\ 1997a), a
version of TreeSPH (\cite{her89}; \cite{kat96}) implemented on
massively parallel supercomputers.  We use $64^3$ dark matter particles
and $64^3$ gas particles, with an additional $2\times 64^3$ neutrino
particles for the CHDM model, evolving within a periodic cube of
11.111$h^{-1}$~comoving Mpc on a side and a gravitational softening
length of 3$h^{-1}$~comoving kpc (equivalent Plummer softening).  We
include a prescription for
converting gas into stars (\cite{kat96}).  The starting redshift and
the mass resolution for the dark matter and gas particles for each
model are listed in Table~\ref{table: cosmo}.  For the LCDM,
TCDM, and OCDM simulations, the initial conditions
are random realizations of a Gaussian field with the appropriate
initial power spectrum; we use the same Fourier phases for each of the
three models to minimize the effect of ``cosmic variance''
in a finite volume. 
The CHDM initial conditions were generated with routines provided by
\cite{kly97b} and thus have different random phases.  
Particles have individual timesteps according to their physical state;
the total number
of smallest timesteps to $z=0$ and the total CPU time required 
are listed in Table~\ref{table: cosmo}.  The calculations
were run on Cray T3Es at the San Diego and Pittsburgh Supercomputing
Centers, typically using 16 processors.

When computing radiative cooling and heating rates, we include the
effects of photoionization by a spatially uniform UV background, 
determining ionic abundances by requiring a balance between creation
and destruction rates for each ionic species (see \cite{kat96}).
We take the shape and intensity of the UV background spectrum $J_\nu$
from the calculations of Haardt \& Madau (1996, hereafter \cite{haa96}),
who assume that the UV background comes predominantly from quasar
light reprocessed by the clumpy IGM.  However, we use the results of
a new version of this calculation (kindly provided by Piero Madau),
which adopts Zheng et al.'s (1997) $\alpha=-1.8$ power law index for the mean
intrinsic quasar spectrum instead of the earlier value of $\alpha=-1.5$.
We will still refer to this $J_\nu$ as the HM spectrum.

While photoionization has little effect on the dynamical evolution 
of the gas in our simulations, it plays a crucial role in our \lya
forest calculations because it determines the neutral hydrogen 
fraction, and hence the \lya opacity, in diffuse intergalactic gas.
The parameter that controls the relation between neutral fraction
and gas overdensity is the HI photoionization rate $\ggh$,
the cross-section weighted integral of $J_\nu$.
Since the overall normalization of $J_\nu$ is not precisely
known from either observation or theory, and since the relation between
neutral fraction and overdensity also depends on the uncertain
parameters $\Omega_b$, $h$, and $\Omega$, we follow the now-standard
approach of renormalizing $J_\nu$ by a constant factor chosen so that
artificial spectra from the simulation reproduce the observed value 
of the mean \lya flux decrement (see, e.g., \cite{mir96}).
For each model, we choose this renormalization factor in order to match
the mean decrement $\bar{D}=0.32$
measured by Rauch et al.\ (1997b) at $z=3$.
We apply the {\it same} renormalization factor at all redshifts, so
the evolution of $\ggh$ (shown in Figure~\ref{fig: matchJ} below)
is precisely that predicted by HM except for this overall multiplicative
constant.  The values of the $J_\nu$ renormalization factors are
listed in Table~\ref{table: cosmo}, and they would not be significantly
different if we chose to match the Rauch et al.\ (1997b) results
at $z=2$ instead of $z=3$.  In practice, we multiply the HI optical
depths by the inverse of this factor instead of recalculating
spectra with an increased $J_\nu$; the two procedures have the same effect on
the highly photoionized gas that produces \lya forest absorption at high
redshifts.  At low redshifts the procedure should also work, albeit not
quite as well owing to a small number of features where
collisional ionization cannot be ignored. 

We extract artificial \lya absorption spectra from the simulations
using the TIPSY package (\cite{kat95}; see \cite{her96} for details).  
We generate 400 spectra along randomly selected lines of
sight (LOS) through the simulation volume at output redshifts of $z=3,
2.5, 2, 1.75, 1.5, 1.25, 1, 0.75, 0.5, 0.25$ and 0.
We sample the spectra with
resolution $\Delta\lambda=0.06$\AA, roughly
corresponding to that of HIRES.  We ``fit a continuum'' to each small
spectral segment (equal in length to the size of the simulation box)
by rescaling all flux values so that the highest flux pixel has
transmission of unity --- specifically, 
we multiply the flux values in all pixels by $\exp(\tau_{\rm min})$,
where $\tau_{\rm min}$ is the optical depth of the highest flux
pixel in the segment.
This procedure removes $\sim 5$\% of the flux at $z\sim 3$
(depending somewhat on cosmology), but for $z\la 2$ the amount of flux
removed is negligible.  After rescaling the continuum, we
add Gaussian random noise corresponding
to a signal-to-noise ratio $S/N=30$.  Varying $S/N$ over a reasonable range,
e.g., between 20 and 50, does not have a significant effect on our
conclusions.  In the future, we will construct artificial spectra more
closely matched to the properties of low-$z$ data; for our current purposes, 
the level of similarity here is sufficient.

\begin{figure*}
\centerline{
\epsfxsize=5.3truein
\epsfbox[65 0 550 740]{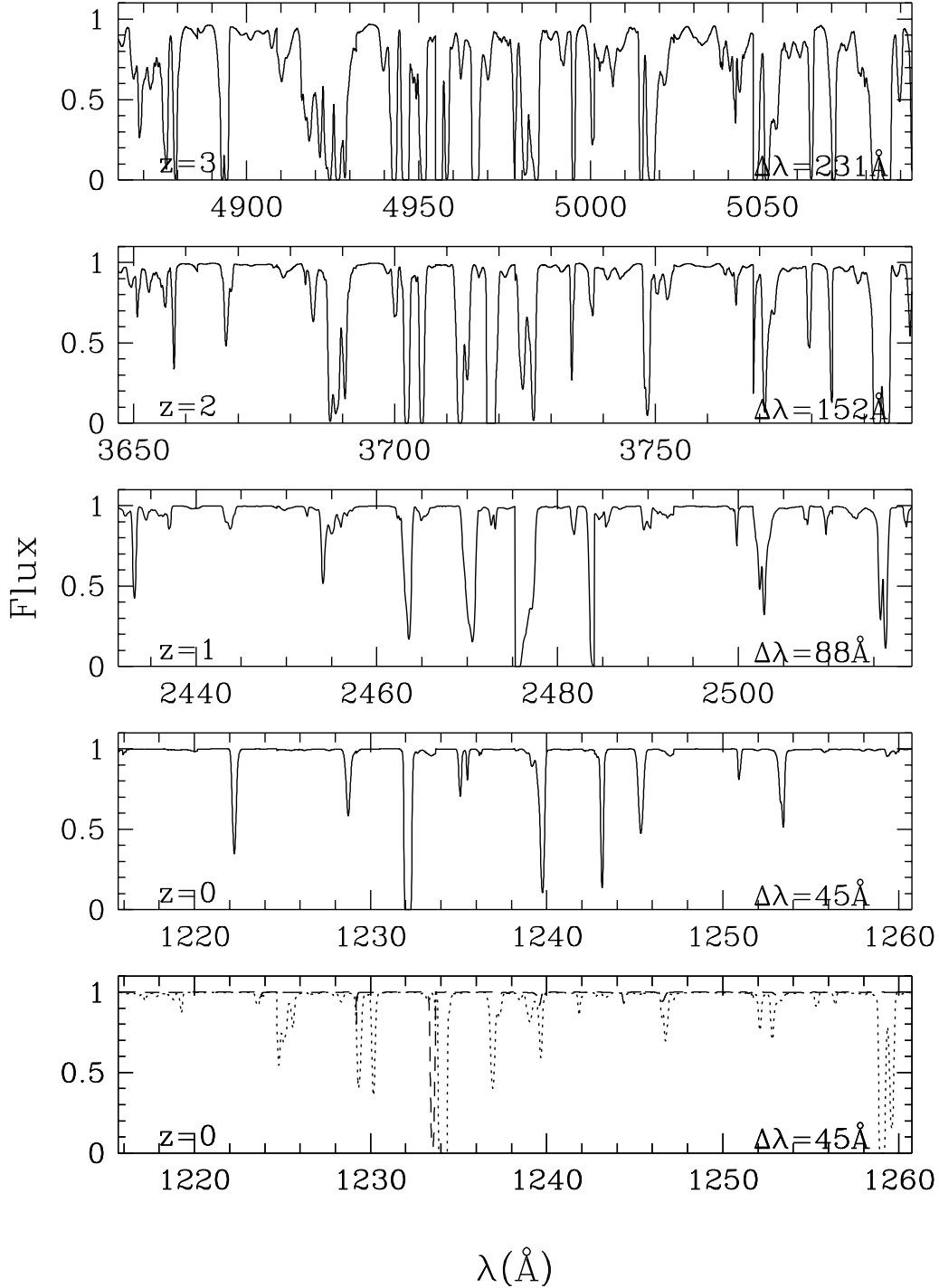}
}
\caption{              
\label{fig: spectra}
Ten artificial spectra at $z=3,2,1$ and 0
from our LCDM simulation, concatenated.  The same lines of sight are
chosen at each redshift.  The total wavelength coverage of the ten
spectra is shown, in \AA, in the lower right.  The strength and number
of absorption features decrease to lower redshift.  The bottom panel
shows the $z=0$ spectra for the structure evolution (dashed line) and
$J_\nu$ evolution (dotted line) scenarios discussed in \S\ref{sec:
ewanal}.
}
\end{figure*}

Figure~\ref{fig: spectra} shows ten sample spectra from the LCDM simulation
at $z=3,2,1$ and 0, concatenated to give a wavelength coverage as
indicated in the lower right corner.  The same lines of sight are shown
at each redshift, so to some extent individual features can be
compared from one spectrum to another, though absorbing structures
may move into or out of the line of sight over time.
The most dramatic evolutionary effect is the paucity of
absorption features at low redshift as compared to high redshift.
The change would be smaller if we plotted the same wavelength
interval instead of the same comoving distance in each panel, but one
can also see that individual features generally become weaker with time
and that the mean flux decrement drops sharply between $z=3$ and $z=1$.
We will discuss the cause of these changes and the significance of
the bottom panel with the dashed and dotted lines in \S\ref{sec: ewanal}.

We fit Voigt profiles to each spectrum using the automated Voigt profile
fitter AutoVP (\cite{dav97b}), yielding a column density, a $b$-parameter
(i.e., a Gaussian width of the optical depth profile), 
and a rest equivalent width for each line.  
\lya absorbers are typically fully resolved at all redshifts
in our artificial spectra, in contrast to currently available HST
spectra.  This represents a systematic difference  that may affect
the comparison of models to data at low redshift, although for most
of the statistics we consider it is unlikely to dominate over other
uncertainties.

\begin{figure*}
\centerline{
\epsfxsize=5.3truein
\epsfbox[65 0 550 740]{fig2.ps}
}
\caption{              
\label{fig: color}
Column density of neutral hydrogen through
a $200\kms$ slice of
the LCDM simulation volume at $z=3,2,1.5,1,0.5,$ and 0
(left-to-right and top-to-bottom).
Red corresponds to  
$10^{13}\cdunits \la \nh \la 10^{14}\cdunits$, 
orange to a column density of $10^{14}\cdunits \la \nh \la 10^{15}
\cdunits$, 
yellow to $ 10^{15}\cdunits \la \nh \la 10^{16}\cdunits$, and      
white to $\nh \gtrsim 10^{16}\cdunits$.         
}
\end{figure*}

Figure~\ref{fig: color} presents a different view of absorption in 
the LCDM simulation, showing the HI column density in a $200\kms$
thick slice at redshifts 3, 2, 1.5, 1, 0.5, and 0.  The dimmest
red level in the plot corresponds to a column density of
$\nh \approx 10^{13}\cdunits$, which in turn corresponds to a
line-center optical depth $\tau_c \approx 0.25$ for a Voigt-profile
line with a $b$-parameter of $30\kms$.
Orange corresponds to a column density of $10^{14}\cdunits \la \nh \la 10^{15}
\cdunits$ and
yellow corresponds to $ 10^{15}\cdunits \la \nh \la 10^{16}\cdunits$.
The saturated, white level corresponds to $\nh \gtrsim 10^{16}\cdunits$,
about a factor of 10 below the column density of a Lyman limit system.
The positions of the compact, saturated dots typically correspond
to the positions of galaxies (clumps of stars and cold, dense gas),
although the luminous regions of the galaxies (and the cross-sections
for producing damped \lya absorption) would be virtually invisible
on the scale of Figure~\ref{fig: color}.

From inspection of Figure~\ref{fig: color}, one can anticipate many
of the conclusions that we will reach from more detailed analyses below.
There is rapid evolution in the number of absorbers above a given
threshold between $z=3$ and $z=1.5$, but
the evolution is much slower between $z=1.5$ and $z=0$.  
The backbone of the structure traced by \lya absorption does not
change radically even over the full range $z=3$ to $z=0$;
the evolution of the forest seems to correspond more to a 
shifting of contour levels rather than a transformation of the underlying
structure.  Finally, for any given choice of threshold, there will be
a much closer correspondence between absorbing regions and galaxies at
low redshift than there is at high redshift.

\section{Evolution of \dndz\ and \dndw}\label{sec: dndz}

\subsection{Observations of \dndz}

At high redshift ($2\la z\la 3$), \cite{kim97} find $\gamma =2.78\pm
0.71$ (where $\gamma$ is defined in equation~[\ref{eqn: gamma}]) from
HIRES spectra.  Ground-based investigations prior to HIRES provided a
somewhat confusing picture, some showing strong evolution
($\gamma\approx 2.8$; Lu et al.\ 1991) and some showing weak evolution
($\gamma\approx 1.7$; \cite{bec94}).  We believe that the \cite{kim97}
determination is more secure and most directly comparable with
simulations, since the HIRES spectra are less subject to line blending
and make it possible to detect \lya absorbers
down to $W_r\sim 0.02$\AA.  However, since the number of quasars currently
observed with HIRES is small, the statistics are poor.  The
amplitude of \dndz\ quoted by \cite{kim97} is for lines with
$\nh>10^{13.77}\cdunits$, while our chosen rest equivalent width limit
($W_r>0.24$\AA) corresponds (for typical $b$-parameters)
to $\nh\ga 10^{13.9}\cdunits$; we
have therefore applied a small correction to \cite{kim97}'s value of $(dN/dz)_0$
based on the overall column density distribution taken from their
paper.

For $z\la 1.5$, \cite{wey98} have published the most complete analysis
of evolution in the Key Project sample, obtaining $\gamma=0.26\pm 0.22$ and
$(dN/dz)_0=32.7\pm 4.2$.  The sample we use for comparison is their uniform
detection limit sample of lines with $W_r>0.24$\AA\ (including \lya systems
with associated metal lines), although
the results from other samples are not markedly different.
\cite{wey98} do find a trend of more rapid evolution of stronger lines,
an issue we will address in \S\ref{sec: dndw} below.

\subsection{Modeling Uncertainties}\label{sec: uncert}

The size of our simulation volume introduces significant uncertainties
when comparing the models with the data.  By $z=1$, all our models have
nonlinear
density fluctuations on the scale of the box length.  Hence, the
artificial spectra at low-$z$ are not necessarily a representative
sample for that cosmology, since the large scale power absent in these
small volume simulations could have a substantial effect on the density
fluctuations at all scales.  We do not yet
have the ability to gauge this systematic
uncertainty, as to do so would require simulating a larger volume than
is currently computationally feasible.  Nevertheless, we can examine
some general trends that should be valid despite these limitations.
Furthermore, we have explicitly minimized the impact of finite volume
statistical fluctuations on comparisons among the 
OCDM, TCDM, and LCDM models by using 
the same initial phases and lines of sight in each simulation.
The differences between these results should therefore reflect
inherent differences among the models (with the caveat that
differences in power on scales larger than the box size could impact
smaller scales).

Another uncertainty arises from the finite resolution of these
simulations.  As Wadsley \& Bond (1997) have noted, the number of intermediate
density \lya forest absorbers ($\nh\ga 10^{15.5}\cdunits$, or $W_r\ga
0.5$\AA) at $z=3$ is somewhat sensitive to numerical resolution, since
these systems are often associated with virialized
``mini-halos'' (\cite{ike86}; \cite{ree86}) of small mass and
spatial extent.  At high redshift, these mini-halos
probably contribute no more than half of the total number 
of absorbers above $W_r\sim 0.6$\AA\ (\cite{gar97a}) and a much smaller
fraction of the number above $W_r \sim 0.24$\AA.  
However, as we shall show in \S\ref{sec: colspa}, 
a given equivalent width absorber corresponds to higher
overdensity gas at lower redshift, meaning that the fractional
contribution of mini-halos above $W_r\sim 0.24$ will increase at lower
redshifts.  If our simulations are unable to resolve these objects, it
is possible that we are increasingly underestimating the total number
of lines at lower redshifts.  We will eventually conduct
higher resolution simulations to examine this possibility in more detail.
Recently Theuns et al.\ (1998b) and Bryan et al.\ (1998) have
examined the sensitivity of {\it high} redshift \lya forest
predictions to the numerical resolution of hydrodynamic simulations.
Both groups find that the predicted column density distribution is
reasonably robust to changes in resolution but that convergence
of the $b$-parameter distribution requires very high resolution.

In addition to these numerical uncertainties, 
there are sizeable uncertainties in the UV background spectrum.
We take the evolution of $J_\nu$ directly from \cite{haa96}, who
calculated the intensity based solely on the observed quasar population
at various epochs.  However, there could be an important contribution
to the metagalactic flux from other sources such as starburst galaxies
and AGNs, especially as the quasar population declines.  This does not
necessarily imply that $J_\nu$ will be higher than \cite{haa96} predict, since
corrections for dust and intervening HI absorption can be large and are
uncertain (\cite{dev98}).  Furthermore, the intrinsic spectral energy
distribution of quasars, assumed by \cite{haa96} to follow a power law
with slope $\alpha=-1.8$, is not well determined, and it could in principle vary
with redshift, particularly through a correlation between luminosity
and its slope (\cite{kor98}).  These variations would be directly
reflected in the number of HI lines observed.  We consider this issue
further in \S\ref{sec: ewcosmo}.

In sum, we expect the simulation predictions of the \lya forest to
be fairly reliable at high redshift but less so at low redshift, where
the effects of the finite simulation volume and finite resolution
are more significant, the UV background history is more uncertain,
and the noise and resolution properties of our artificial spectra
are less closely matched to current observational data (see \S\ref{sec: sims}).
We are not yet in a position to gauge the magnitude of some of these
uncertainties, especially those related to purely numerical effects.
We will therefore not attempt to rule out any of the cosmological
models that we consider on the basis of low-$z$ \lya forest data.
Instead, we focus in this paper on the physical processes that 
give rise to \lya forest absorption at various redshifts, with particular
attention to those trends that are independent of the details of the
cosmological model.

\subsection{The Evolution of \dndz\ in Various Cosmologies}\label{sec: ewcosmo}

\begin{figure*}
\centerline{
\epsfxsize=4.0truein
\epsfbox[95 365 465 735]{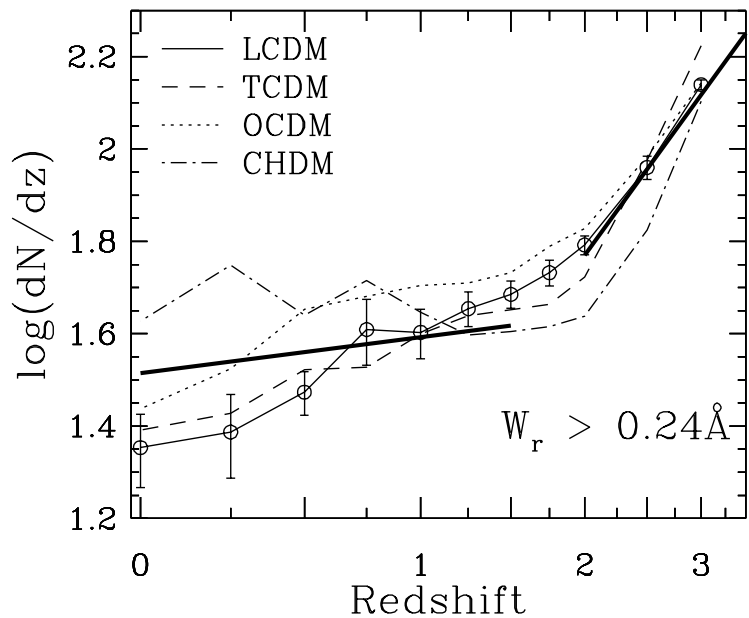}
}
\caption{              
\label{fig: ewcosmo}
Evolution of 
\dndz\ in various cosmologies.  The thick
solid line above $z\geq 2$ is from the observations of \cite{kim97},
while the thick solid line at $z\leq 1.5$ is a fit from
\cite{wey98}.  All cosmologies show a change in evolution of \dndz\
after $z\sim 2$.  
The horizontal axis scale in this and subsequent evolution figures
is linear in $\log(1+z)$, so that power laws in $(1+z)$ appear
as straight lines.
}
\end{figure*}

Figure~\ref{fig: ewcosmo} shows the redshift evolution of
\dndz\ in our four cosmologies from $z=3$ to $z=0$.  
For the LCDM model, we show $1\sigma$ statistical uncertainties
computed by taking the dispersion among 10 
subsamples of 40 random LOS and dividing by $\sqrt{N-1}=3$.
Statistical errors for the other models are comparable.
Note, however, that these error bars do not incorporate systematic
uncertainties from numerical limitations or mismatch between
observational and theoretical analysis procedures, and they 
underestimate the true statistical error because we have only a
single realization of structure for each cosmological model.
Table~\ref{table: dndz} shows $\gamma$ inferred from the
observations and the simulations in various redshift intervals.  We
have not listed uncertainties for $\gamma$ from the simulations, since
at high redshift they are smaller than the uncertainties from \cite{kim97},
and at low redshift they are likely to be dominated by unquantified
systematic errors.

For $2\la z\la 3$, the LCDM and OCDM models are in good agreement with
the data, while the TCDM and CHDM models show steeper
evolution, with values of $\gamma$ that lie outside the formal $1\sigma$
error of the \cite{kim97} estimate.
The number of lines with $W_r>0.24$\AA\ in the CHDM model at $z=2$ is
significantly lower than observed, a failing that reflects the
paucity of small-scale structure at early epochs in a CHDM universe.
The intensity of the ionizing background has been normalized in each model
to reproduce the Rauch et al.\ (1997b) measurement of the mean \lya flux
decrement at $z=3$, so the value of \dndz\ cannot be altered by changing
$J_\nu$ without spoiling the agreement with other data.
Detailed comparisons of these models to
high-redshift data with a more careful treatment of systematic
uncertainties will be presented elsewhere.

At $z\sim 2$, the rate of evolution of \dndz\ starts to decrease in
all the models.  The history of structure formation is very different
from one model to another, and the similarity of the \dndz\ curves
is an important clue that the break in \dndz\ evolution is not
caused primarily by the gravitational growth of structure 
(see \S\ref{sec: ewanal} below).
For LCDM, TCDM, and OCDM, \dndz\ continues to drop with redshift
down to $z=0$, while for CHDM the
slope is consistent with zero below $z\sim 1.5$.  The spectral
resolution of our artificial spectra is considerably higher than
that typical of FOS spectra; 
blending caused by poorer resolution tends to raise
the number of lines above $W_r>0.24$\AA, but the overall rate of
evolution does not change significantly.  At face value, the CHDM model
is in better agreement with the low redshift data than LCDM, TCDM or OCDM.
However, uncertainties in $J_\nu$ at low redshift are much too
large to allow any meaningful discrimination between cosmologies based
on \dndz\ evolution, even disregarding the other uncertainties noted
in \S\ref{sec: uncert}.

To gauge the sensitivity of $\gamma$ to the evolution of
$J_\nu$, we determine the evolution of the photoionization rate $\ggh$
for the LCDM simulation 
that would be required to produce perfect agreement with the
observed evolution of \dndz\ at $z\la 1.5$.  
We obtain an initial guess for the
required evolution of $\ggh$ by simply multiplying the \cite{haa96} value of
$\ggh$ at redshifts $z=1.5,1,0.5$, and 0 by the ratio of the observed
\dndz\ and \dndz\ from the LCDM model with the original \cite{haa96}
$\ggh$.  This first correction brings \dndz\ within 10\% of the
observed value.  Minor adjustments then produce the evolution of $\ggh$
shown in Figure~\ref{fig:  matchJ}.  The solid line is the evolution of
$\ggh$ directly from \cite{haa96}, while the dashed line shows the
evolution of $\ggh$ that would yield exact agreement with the
evolution of \dndz\ from \cite{wey98}, for the LCDM model.  
The differences are not large --- at most 50\% at $z=0$ --- and other
cosmologies would have differences that are typically smaller than that
of the LCDM model (cf. Figure~\ref{fig:  ewcosmo}).  An independent
theoretical estimate of the evolution of $\ggh$ from 
Fardal, Giroux, \& Shull (1998, their Q1 model)
is shown as the dotted line.  The agreement between the \cite{haa96}
and Fardal et al.\ (1998, hereafter \cite{far98}) 
models is impressively good, but
both of them rest on the assumption that the UV background is
produced entirely by quasar sources, and both of them rely
on Pei's (1995) estimate of the evolution of the quasar luminosity function.

\begin{figure*}
\centerline{
\epsfxsize=4.0truein
\epsfbox[95 365 465 735]{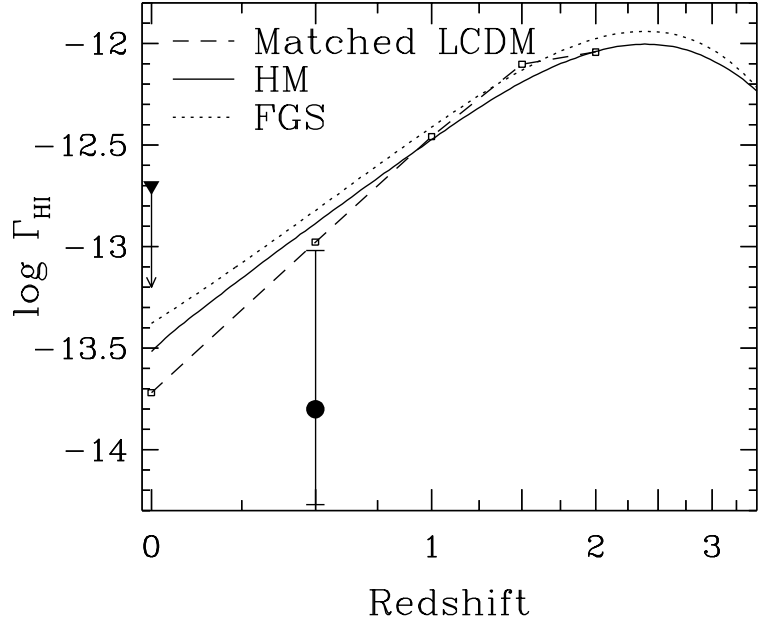}
}
\caption{              
\label{fig: matchJ}
The evolution of $\ggh$ from \cite{haa96}
(solid line) and \cite{far98} (dotted line), and the evolution of
$\ggh$ required to force agreement between the LCDM model and the
observed low-$z$ \dndz\ (dashed line).  The differences are
small compared to the uncertainties in observational estimates of $\ggh$ from
the proximity effect (\cite{kul93}, solid circle) and from H$\alpha$
flux measurements (\cite{vog95}, solid triangle).
}
\end{figure*}

Vogel et al.\ (1995) have used H$\alpha$ surface brightness
measurements of a local HI cloud  to place an upper limit on
$\ggh$ at $z=0$, which is shown by the triangle in Figure~\ref{fig: matchJ}.
The circle at $z=0.5$ shows Kulkarni \& Fall's (1993)
estimate of $J_\nu$ from the proximity effect 
(converted using $\ggh=2.64\times 10^9 J_\nu$ in cgs units; see
\cite{far98}).
The uncertainties in these measurements are much larger than the
variation in $\ggh$ required to reconcile the LCDM model with the Key
Project data.  
Much more precise observational determinations of $\ggh$ would
be needed to directly test cosmological models using
the value of \dndz\ at low redshift.

\begin{figure*}
\centerline{
\epsfxsize=4.0truein
\epsfbox[95 365 465 735]{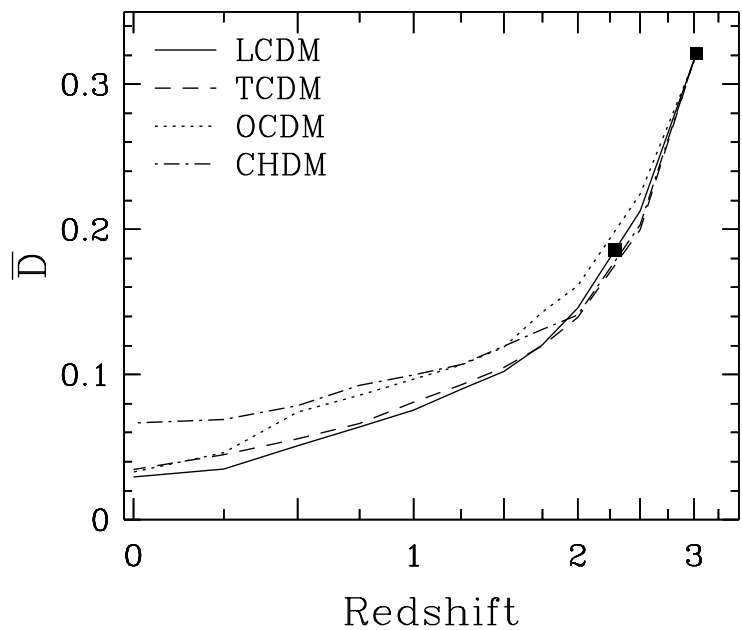}
}
\caption{              
\label{fig: tauevol}
Evolution of the mean \lya flux decrement,
$\bar{D} \equiv \langle 1-e^{-\tau}\rangle.$
Filled squares are $\bar{D}$ from sample of
Keck/HIRES spectra analyzed by Rauch et al.\ (1997b).  All models agree
well with $\bar{D}$ evolution for $z = 2-3$.
}
\end{figure*}

A more promising route to testing cosmological models is to choose
the value of $\ggh$ at each redshift in order to match one observational
datum, then use other statistical properties of the \lya forest
as tests of the model.  One option is to adopt \dndz\ as the
``normalizing'' observation -- in this case we would use the
values of $\ggh$ shown in Figure~\ref{fig: matchJ} for the LCDM model.
An alternative normalizing datum is the mean flux decrement $\bar{D}$,
which has the advantages of not depending on a line identification 
algorithm and of being independent of spectral resolution.
Figure~\ref{fig: tauevol} shows the evolution of $\bar{D}$ in our
simulations, with data points at $z=2.29$ and $z=3.02$ from
Rauch et al.\ (1997b).  As discussed in \S\ref{sec: sims}, we
adjust the intensity of $J_\nu$ (with the
\cite{haa96} shape) in each model to obtain agreement
with the observed $\bar{D}$ at $z=3$.  
The subsequent predicted evolution of $\bar{D}$ relies on
the \cite{haa96} UV background history.
With this history, all of the models continue to match the
Rauch et al.\ (1997b) data at $z=2.29$.
We do not know of any determinations of $\bar{D}$ from HST spectra
at low redshift,
although it should be straightforward in principle to estimate
from existing data.  
The trends in the evolution of $\bar{D}$ are similar to the trends
in the evolution of \dndz, as one would expect.

\subsection{The Cause of the Break in \dndz\ Evolution}\label{sec: ewanal}

The primary driver of evolution in \dndz\ at high redshift is
the expansion of the universe (\cite{her96}; \cite{mir96}; \cite{bi97}).
The {\it physical} density associated with a given {\it overdensity}
drops as the universe expands, and as a result the neutral hydrogen
fraction, which is proportional to the recombination rate, also
drops.  The neutral column density of any structure that is expanding with
residual Hubble flow drops because of the reduced neutral fraction and
because of expansion along directions perpendicular to the line of sight.
At any redshift there are fewer high column density systems than
low column density systems, so as the expansion of the universe
drives the column density of each system down, the number of lines above
any given threshold column density declines.

It is clear from observations and from our simulations that the
evolution of \dndz\ is much slower at low redshift, with a 
rather sharp transition in the evolution rate occurring at $z \sim 1.7$.
We consider two physical processes that could influence \dndz\ 
evolution in addition to the Hubble expansion effects described above:
\begin{enumerate}
\item ``Structure evolution": gravitational growth of structure drives
gas from low density regions into filaments and sheets and from filaments
and sheets into collapsed objects, so that the physical structures
producing \lya forest absorption are fundamentally different at
low and high redshifts.
\item ``$J_\nu$ evolution'': the intensity of the UV background drops
at $z<2$ because of the declining quasar population, so that the lower
recombination rate at low redshifts is countered by a lower photoionization
rate.
\end{enumerate}

Clearly both of these effects must operate in a complete model, but
we would like to know whether one predominates in changing the evolution rate.
To investigate this question, we consider idealized treatments of the LCDM 
simulation that isolate the two effects.  The ``structure evolution 
scenario'' is
represented by a model in which the intensity of the \cite{haa96}
ionizing background is held fixed at its value at $z=2$ and only the
distribution of matter is varied from $z=3$ to $z=0$.  The ``$J_\nu$
evolution scenario'' is represented by a model in which the gas 
distribution is held fixed in comoving coordinates
at its $z=2$ distribution but the ionizing
background is varied with redshift as prescribed by \cite{haa96}.
Hubble expansion is included in both scenarios.  
In a pure $J_\nu$ evolution scenario, the changes in 
Figure~\ref{fig: color} would be exactly equivalent to a shift
in the color map from one redshift to another.  Figure~\ref{fig: contour}
presents the evolution of the projected column density in this
$200\kms$ slice (the same one as Figure~\ref{fig: color}) in the form
of a contour plot.  Contours are spaced by factors of 10 in column
density, with the heavy contours corresponding to column densities
of $10^{11}$, $10^{14}$, and $10^{17}\cdunits$.
A factor of ten change in $J_\nu$ at fixed redshift would correspond
to a shift of one contour level in this plot, while for a fixed
$J_\nu$ and fixed comoving gas distribution, the column densities
would be proportional to $(1+z)^5$ (see discussion below).

\begin{figure*}
\centerline{
\epsfxsize=5.3truein
\epsfbox[65 0 550 740]{fig6.ps}
}
\caption{              
\label{fig: contour}
Neutral hydrogen column density in the same $200\kms$ thick slice as that
shown in
Figure ~\ref{fig: color}.  The contours are spaced by factors of 10
in column density. The heavy contours correspond to column densities   
of $10^{11}$, $10^{14}$, and $10^{17}\cdunits$.
}
\end{figure*}

\begin{figure*}
\centerline{
\epsfxsize=4.0truein
\epsfbox[95 365 465 735]{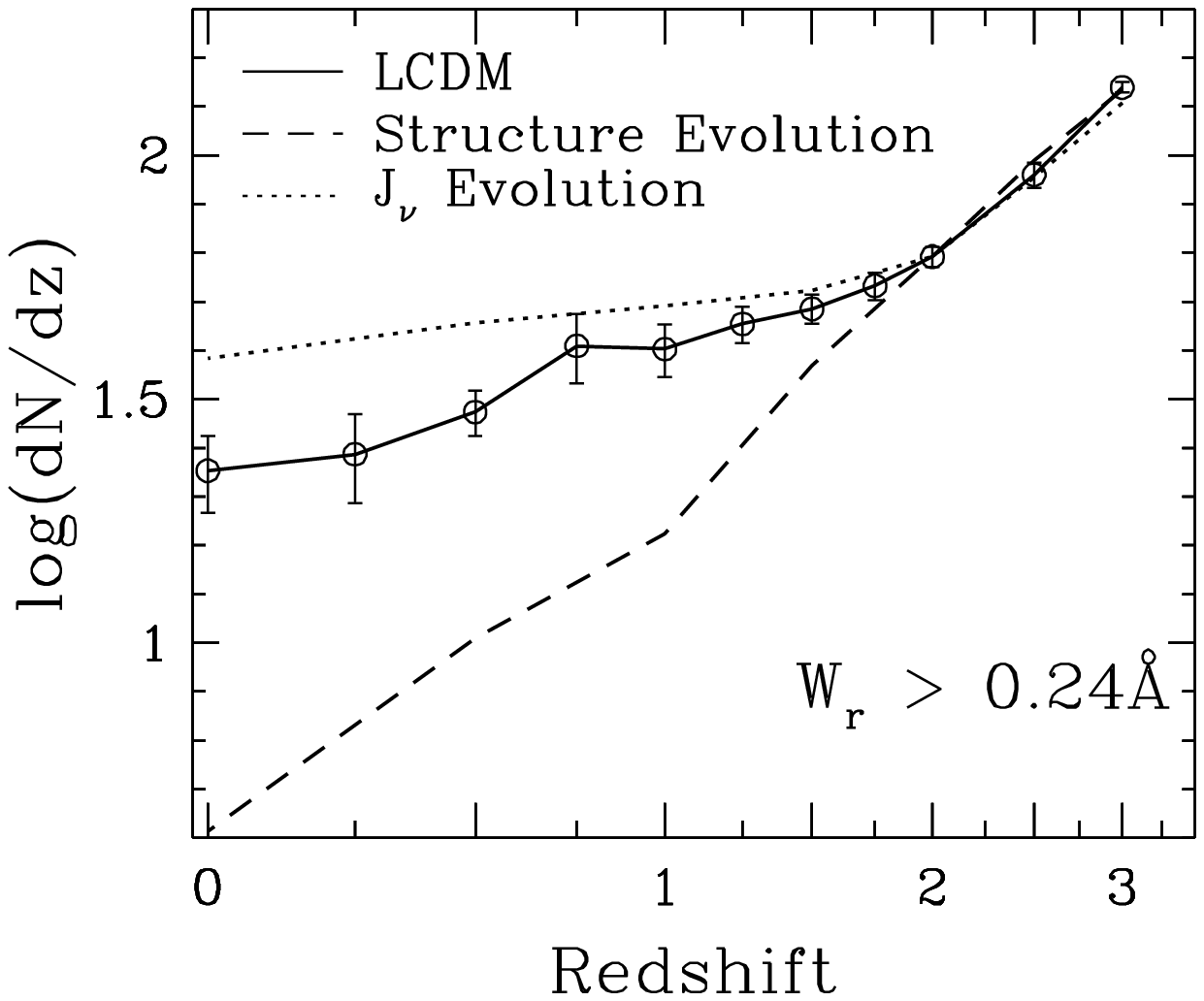}
}
\caption{              
\label{fig: ewanal}
\dndz\ from varying only structure or only
$J_\nu$.  For $2\la z\la 3$, the dominant effect is Hubble expansion,
as $J_\nu$ and structure both remain roughly constant.  For $z\la 2$,
the structure evolution (dashed line)
and $J_\nu$ evolution (dotted line) models diverge rapidly, demonstrating
that the observed break is caused predominantly by changes
in $J_\nu$.  However, \dndz\ from the $J_\nu$ evolution model evolves
more slowly than that from the full LCDM simulation (solid line), 
indicating that structure formation plays a non-negligible role.
}
\end{figure*}

The bottom panel of Figure~\ref{fig: spectra} shows artificial spectra at $z=0$
along the same ten lines of sight shown in
the panel above, from the structure evolution scenario (dashed
line) and the $J_\nu$ evolution scenario (dotted line).
Comparing to the true $z=0$ spectrum, it is evident that,
while neither idealization exactly reproduces the full evolution,
the $J_\nu$ evolution scenario comes much closer.
Figure~\ref{fig: ewanal} quantifies this impression in terms of \dndz.
The solid line shows the true \dndz\ evolution of the LCDM simulation,
reproduced from Figure~\ref{fig: ewcosmo}, and the heavy dashed
and dotted lines show \dndz\ from the structure and $J_\nu$ evolution
scenarios, respectively.  The structure evolution scenario 
predicts a redshift evolution with no apparent break;
the power law from $z\ga 2$ continues smoothly down to lower
redshifts, with $\gamma\approx 2.5$ from $z=2$ to $z=0$.  Conversely,
the $J_\nu$ evolution scenario predicts a distinct break
beginning around $z\sim 2$.  We conclude that the break in \dndz\ 
evolution in the simulations (and, by implication, in the observations)
is caused primarily by the decline of the UV background at low redshift.
Coincidentally, the evolution of the \cite{haa96} background is just 
enough to roughly cancel the effects of Hubble expansion,
yielding $\gamma\approx 0$.  A similar analysis for the TCDM or CHDM models 
yields the same general conclusion.

Since the dotted and solid curves in Figure~\ref{fig: ewanal} do not
agree perfectly, it appears that structure evolution does have
some influence on \dndz\ evolution, as one would expect.  The effect
of gravitational growth of the underlying structure is to {\it reduce}
the number of absorbers observed above an equivalent width threshold,
as matter empties from large scale structure into
collapsed objects that have smaller cross-sections of absorbing gas.

The $J_\nu$ evolution scenario for \dndz\ can be approximated analytically.
The simplifying assumption of this scenario is that the gas distribution
remains fixed in comoving coordinates, so that the gas density at 
a comoving position drops as $(1+z)^3$ in proportion to the mean
density of the universe.  If we ignore collisional ionization, then
the neutral hydrogen fraction is proportional to the recombination
rate divided by the photoionization rate, and hence to $(1+z)^3\ggh^{-1}(z)$.
%Assuming that each identified line truly corresponds to an absorber,
%something not necessarily born out by the simulations then 
The neutral hydrogen column density of a given line is equal
to the total column density times the neutral fraction and is therefore
proportional to $(1+z)^5\ggh^{-1}(z)$.  The additional factor of
$(1+z)^2$ accounts for the decrease in total column density caused
by the expansion of the absorber along the two dimensions perpendicular
to the line of sight.  Unless the photoionization rate falls faster
than $(1+z)^5$, the neutral column density of every absorber falls,
and systems that were once above a column density threshold drop below it as
the universe expands, driving \dndz\ down.  A compensating effect is that
the redshift interval $\Delta z$ that corresponds to a comoving distance
interval $\Delta X$ is proportional to the Hubble parameter $H(z)$. Therefore,
for a fixed comoving structure, the number of lines per unit redshift is
also proportional to $H^{-1}(z)$.

To obtain a quantitative prediction from this approximation,
suppose that at a fiducial redshift $z_f$ the distribution of HI
column densities is a power law,
\begin{equation}
f(\nh) \equiv {dN \over d\nh} \propto \nh^{-\beta}.
\label{eqn: fn}
\end{equation}
Observations imply $\beta \approx 1.5-1.7$ at $z \sim 2-3$
(e.g., \cite{pet93}; \cite{kim97}), and our numerical simulations
yield column density distributions that are reasonably close to power
laws with similar values of $\beta$.  The number of lines above a 
limiting column density is the integral of this distribution,
$F(\nhl) \propto \nhl^{1-\beta}$.
To determine the number of lines above $\nhl$ at some other 
redshift $z$, we can count the lines at redshift $z_f$ above a
different limiting column density
\begin{equation}
\nhl^\prime = \nhl \left({1+z_f \over 1+z}\right)^5 {\ggh(z) \over \ggh(z_f)}.
\label{eqn: nhl}
\end{equation}
Combining this shift in threshold with $F(\nhl)$ and the comoving distance
to redshift conversion, we obtain
\begin{equation}
\left(dN \over dz\right)_{>W_{r,{\rm lim}}} = 
C \left[(1+z)^5 \ggh^{-1}(z)\right]^{\beta-1} H^{-1}(z),
\label{eqn: ewanal}
\end{equation}
where $C$ is a constant chosen to match \dndz\ at $z=z_f$.

In equation~(\ref{eqn: ewanal}) we implicitly assume that an equivalent
width threshold can be identified with a column density threshold.
This identification is exact in the case of optically thin lines,
and it is a reasonable approximation for mildly saturated lines
because the distribution of column densities is much broader than
the distribution of $b$-parameters.  It is clear from 
equation~(\ref{eqn: ewanal}) that \dndz\ should drop rapidly with
redshift if $\ggh$ is constant but that a declining $\ggh(z)$
can cancel the effects of Hubble expansion on \dndz.
This analytic approach provides an
approximate match to the numerical results for fixed comoving structure,
demonstrating the simplicity of the underlying evolution mechanism:
at different redshifts, a different column density or equivalent
width threshold selects a different subset of the line population.
However, the analytic description implicitly assumes that Voigt-profile
decomposition measures correct neutral column densities for a discrete
set of absorbing structures and that the distribution of these column
densities is well described by a single power law; detailed analysis
of the simulations often undermines both of these assumptions.

Our conclusion that the drop in the UV background is the primary
cause of the break in \dndz\ is the same as that reached by
Theuns et al.\ (1998a) on the basis of similar SPH simulations.
We believe that this is also the explanation for the change in
evolution rate found by Riediger et al.\ (1998) in their
pseudo-hydro N-body simulations, although Riediger et al.\ do not
emphasize this interpretation.    

\subsection{The Evolution of \dndw}\label{sec: dndw}

Observations show that the distribution of rest equivalent widths $W_r$
of \lya absorbers is roughly exponential, \ie
\begin{equation}\label{eqn: dndw}
{dN\over dW_r}\propto e^{-W_r/W_r^\star}.
\end{equation}
{}$W_r^\star$ has been determined for high-$z$ \lya absorbers by
several pre-HIRES studies, ranging from $W_r^\star\approx
0.36$\AA\ (\cite{sar80}) to $W_r^\star\approx 0.28$\AA\ (\cite{mur86})
to $W_r^\star\approx 0.16$\AA\ (\cite{car84}).  The differences are
likely attributable to spectral resolution and blending effects
(\cite{car84}).  HIRES studies have focused on the distributions of
$\nh$ and $b$, but given the good agreement of these distributions
between simulations and data at high redshift (e.g., Zhang et al.\ 1995;
\cite{dav97b}), we expect that our high-redshift determinations of
$W_r^\star$ will be in good agreement with the observations as well.  Low-$z$
\lya absorbers are observed to have somewhat lower $W_r^\star$,
although it is unclear whether this difference
reflects a change in the 
intrinsic properties of absorbers or biases introduced by poorer
resolution at low redshift and line blending at high redshift (Lu et al.\ 1991;
\cite{bah96}).

\begin{figure*}
\centerline{
\epsfxsize=4.0truein
\epsfbox[95 365 465 735]{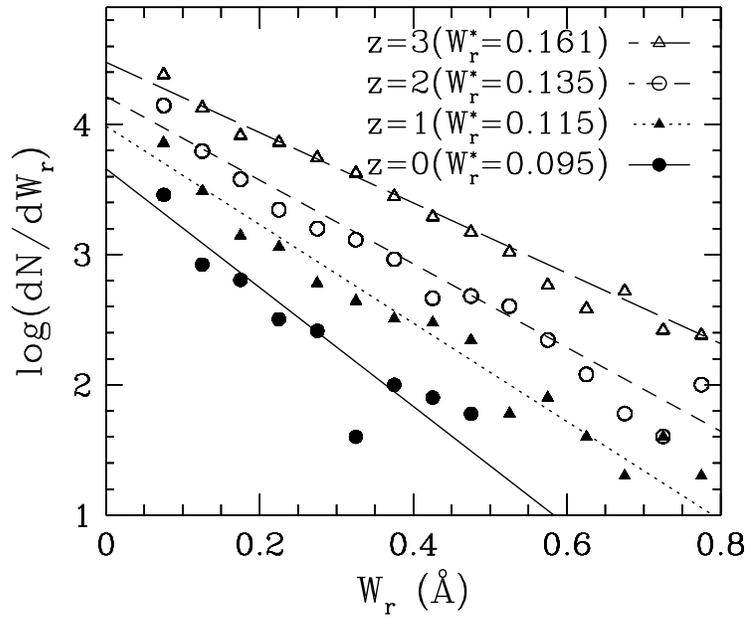}
}
\caption{              
\label{fig: ewdist}
Equivalent width distributions at $z=3,2,1,0$
in the LCDM model.  The best-fit value of $W_r^\star$ 
(eq.~[\ref{eqn: dndw}]) at each redshift is shown in the legend.
The redshift dependence can be roughly parameterized as indicated in
equation~(\ref{eqn: wstar}).  
}
\end{figure*}

Figure~\ref{fig: ewdist} shows the distribution of equivalent widths at
$z=3,2,1,$ and 0 from our LCDM simulation.   The lines show the best fit to
equation~(\ref{eqn: dndw}), and the corresponding values for $W_r^\star$ are
indicated in the legend.  The distributions are truncated at high $W_r$
when the number of absorbers in a bin of width 0.04\AA\ is zero.

The equivalent width distributions from our simulations are reasonably
well described by exponentials at all redshifts.  $W_r^\star$ decreases
steadily and significantly with redshift, from $\approx 0.17$\AA\ at
$z=3$ to $\approx 0.1$\AA\ at $z=0$.  The redshift dependence of
$W_r^\star$ in our LCDM model can be crudely characterized as
\begin{equation}\label{eqn: wstar}
W_r^\star \sim 0.1\times (1+z)^{0.1}.
\end{equation}
{}The other models produce a roughly similar dependence, so this result
is fairly insensitive to the cosmology.  However, it does depend on the high
spectral resolution adopted for our artificial spectra.

To investigate the spectral resolution issue we generated
artificial spectra at $z=0$ with a resolution of $230\kms$ (comparable to
FOS). These lower resolution spectra are well fit by an exponential
distribution with a larger value of
$W_r^\star\approx 0.26$ (except with a significant turnover
for $W_r\la 0.1$\AA\ caused by incompleteness).  $W_r^\star\approx 0.26$
is in reasonable
agreement with \cite{wey98}, who find $W_r^\star=0.283\pm 0.012$.
Thus, the discrepancy between equation~(\ref{eqn: wstar}) and the
(larger) values of $W_r^\star$ deduced from the observations apparently reflects
the influence of line blending rather than a fundamental failing
of the cosmological models.  A direct test of the simulations on
the basis of the equivalent width distribution will require a closer
match of resolution and noise properties between artificial 
and observed spectra, which we defer to future work.

The steepening of $dN/dW_r$ implies that strong \lya absorbers evolve
more rapidly than weak ones.  This trend is consistent with
\cite{wey98}, who find that $\gamma$ is larger for higher equivalent
width absorbers.  Using equation (\ref{eqn: wstar}), we predict the values
of $\gamma$ for the redshift range $z=1.5$ to $z=0$ in different equivalent
width regimes\footnote{We cannot determine this relationship directly from our
spectra up to the highest values of $W_r$ explored by \cite{wey98}
because we have very poor statistics for lines with large $W_r$.}, analogous
to Figure~7 of \cite{wey98}.  The results are shown in Figure~\ref{fig:
ewgamma}.  While the trend from artificial spectra (open circles) is
qualitatively similar to observations (filled squares with error bars),
the data appear to show a more dramatic decrease in $\gamma$ for
low-$W_r$ lines.  Note that the simulation results are derived directly from
equation~(\ref{eqn: wstar}), so the effects of spectral resolution (i.e.,
blending) and lack of sensitivity to small equivalent widths (i.e.,
incompleteness) are probably significant.  Hence it is unclear whether
this represents a failing of the model or systematic differences
between the artificial spectra and the FOS data.  In any case, the 
statistical significance of the discrepancy is barely $1\sigma$.

\begin{figure*}
\centerline{
\epsfxsize=4.0truein
\epsfbox[95 365 465 735]{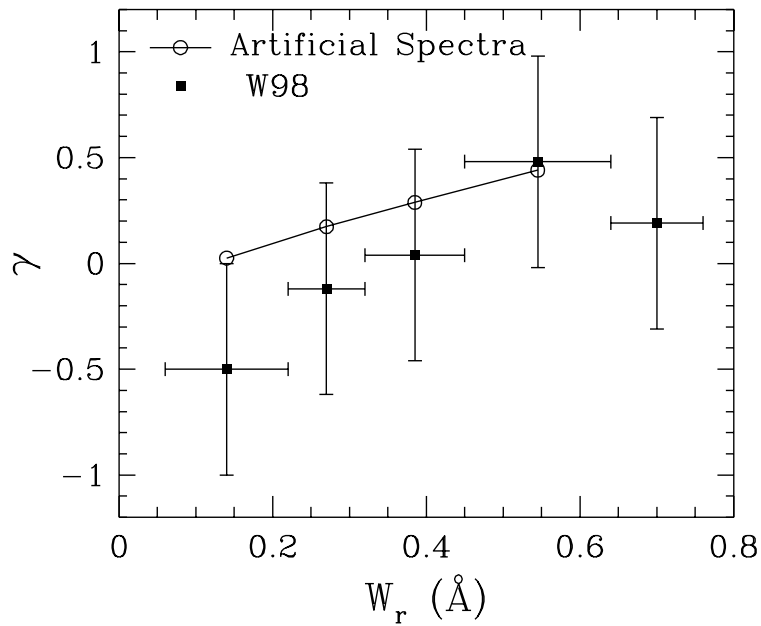}
}
\caption{              
\label{fig: ewgamma}
Equivalent width vs. $\gamma$ (defined in
eq.~[\ref{eqn: gamma}]), as predicted from equation~(\ref{eqn:
wstar}).  The observations of \cite{wey98} are shown as solid squares
with error bars.  The value of
$\gamma$ is larger for strong absorbers, indicating
that they evolve more rapidly than weak absorbers. 
More sensitive observations at low-$z$ should reveal a large population of
weak lines.
}
\end{figure*}

There is further observational evidence that the equivalent width distribution
steepens at lower redshift.  The column density slope found for the
$\langle z \rangle=1.9$ spectrum of Q1331+170 (\cite{kul96}) is steeper 
than that typically found at $z\sim 3$ (\eg \cite{kim97}).  At even larger 
$W_r$, Lyman limit systems (LLS) are seen to evolve more rapidly than \lya
forest absorbers at low redshifts ($\gamma_{\rm LLS} = 1.5\pm 0.39$;
\cite{ste95}).  It is less straightforward to count LLS in our 
simulations because the effects of neutral hydrogen self-shielding
and the contribution from low mass objects must be taken into account 
(e.g., \cite{gar97a}); 
nevertheless, we are planning a careful examination of
their redshift evolution in the future.

The $J_\nu$ evolution scenario, which approximately describes the evolution
of \dndz, predicts the same value of $\gamma$ for all equivalent width
limits.  The trend of steeper evolution for stronger \lya absorbers
probably reflects the physics of gravitational structure formation.
Regions of higher density expand more slowly than regions of lower density
owing to their stronger self-gravity.  As
we shall show in \S\ref{sec: colspa}, lower density gas gives rise to
weaker \lya absorbers.  Thus the fractional cross-section of weaker
\lya absorption increases with time relative to that of stronger \lya
absorption, giving rise to preferentially more weak absorbers at
low-$z$. 
%The overall evolution of the absorbing cross-section is
%reflected in Figure~\ref{fig: color}, where the high-column regions become
%progressively smaller,
%as well as 
%in Figure~\ref{fig: ewanal}, where the $J_\nu$ evolution
%model (without structure evolution) predicts progressively more
%absorbers above $W_r>0.24$\AA\ than the LCDM model
%including structure formation.  
%One can see this more quantitatively in Figure~\ref{fig: cont}, where
%we show contours plots of the same $200\kms$ thick slice as that shown in 
%Figure ~\ref{fig: color}.  The contours are spaced by factors of 10     
%in column density. The heavy contours correspond to column densities     
%of $10^{11}$, $10^{14}$, and $10^{17}\cdunits$.
%Thus, 
While structure formation plays a
subordinate role to the ionizing background in the evolution of \dndz,
it is central to the evolution of $W_r^\star$.

Equation~(\ref{eqn: wstar}) implies that at low redshift, the \lya forest
should be even more dominated by weaker absorbers than it is at high
redshifts.  Using GHRS data, \cite{tri98} estimate \dndz$=102 \pm 16$
for lines with $W_r > 0.05$\AA\ and $0<z<0.28$, about three times
higher than the value that \cite{wey98} find for $W_r > 0.24$\AA.
Equation~(\ref{eqn: wstar}) predicts a ratio of $e^{(0.24-0.05)/0.1}=6.7$,
which suggests that the equivalent width distribution predicted by the
simulations is excessively steep.  Assessment of this possible discrepancy
will require more detailed consideration of spectral resolution effects.
Higher resolution data from STIS should eventually yield more accurate
measurements of the population of weak lines.

\section{Physical Evolution of the Absorbing Gas}\label{sec: physprop}

\subsection{Column Density vs. Overdensity}\label{sec: colspa}

One of the most illuminating results to emerge from hydrodynamic
simulations of the high-redshift \lya forest is the existence of
a tight correlation between the HI column density of an absorption
feature and the average or peak gas density of the structure that
produces it (see, e.g., Zhang et al.\ 1995; \cite{mir96}; \cite{hel98}).  
In principle, the column density of an absorber can be affected
by its density and temperature (which together with $\ggh$ determine
the neutral fraction and neutral gas density) and by its line-of-sight
spatial extent (since the product of spatial extent and neutral gas density
yields the neutral column density).  However, the competition between
photoionization heating and adiabatic cooling leads to a tight correlation
between temperature and density for the unshocked gas that produces most
of the high-z \lya forest absorption (\cite{hg97}), thus linking
two of the three parameters.  Variations in the line-of-sight spatial
extent remain a source of scatter, but they are small compared to the
range of neutral gas densities.

\begin{figure*}
\centerline{
\epsfxsize=5.3truein
\epsfbox[65 200 550 740]{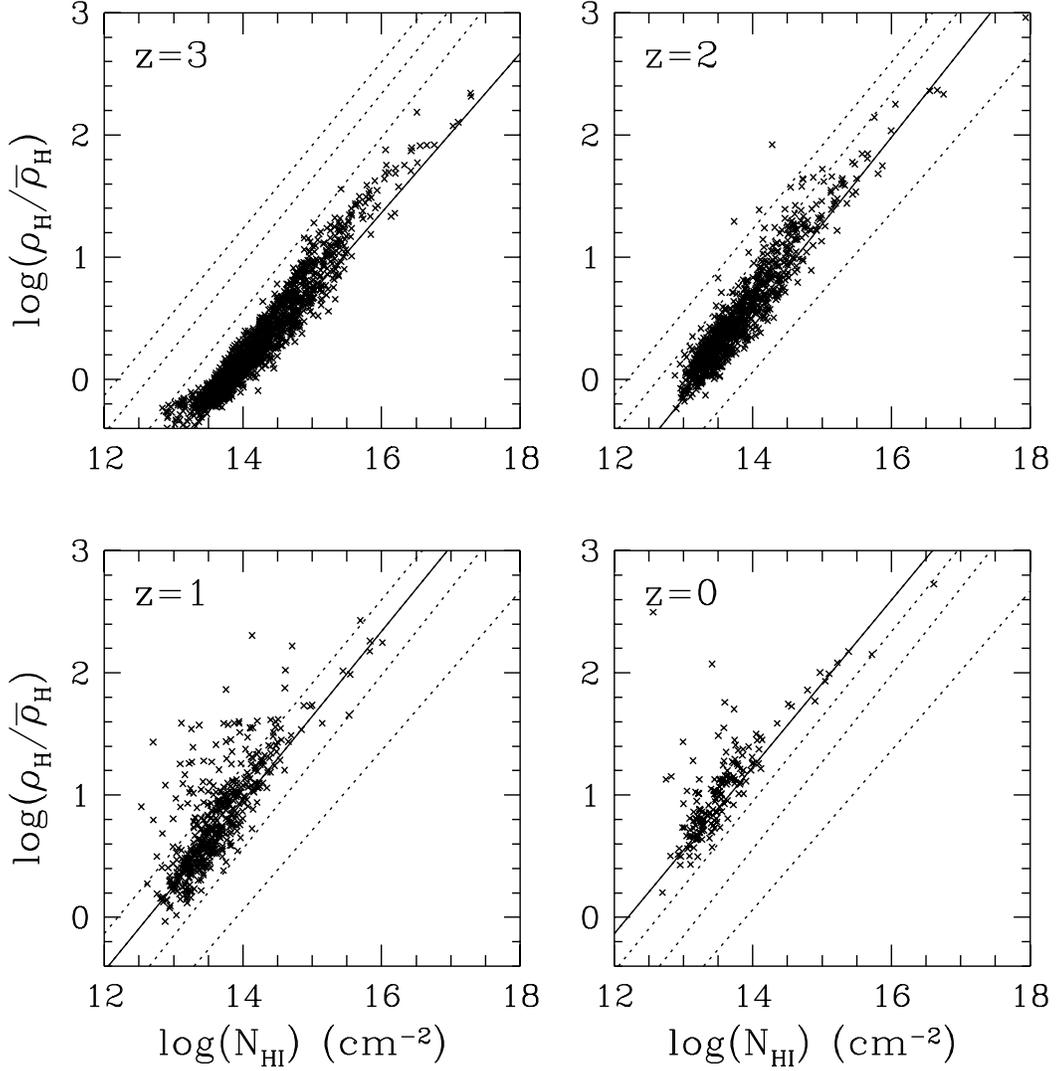}
}
\caption{              
\label{fig: colspa}
The relation between gas overdensity and
column density for the LCDM model, at $z=3,2,1,$ and 0.  The physics
of the photoionized IGM leads to a power law relation that has
roughly constant index, but the overdensity associated with 
a given column density increases with decreasing $z$.  The
solid line in each panel shows the best-fit power law at each redshift,
while the dotted lines show the fits from other redshifts.  Collapsed
objects such as galaxies form at overdensities $\ga 100$, implying
that low-$z$ absorbers will be more correlated with galaxies at a given
}
\end{figure*}

Figure~\ref{fig: colspa} plots the overdensity $\rho_H/\bar{\rho}_H$
against neutral column density $\nh$ for absorbers at
$z=3,2,1,$ and 0 in the LCDM model.
For purposes of this figure, we identify absorbers directly from the simulation
output (without noise or continuum fitting) using the method of 
Hellsten et al.\ (1998): lines are identified as peaks in the optical depth, 
and the column density is summed outwards from the peak 
until a local maximum is reached. 

There is a clear correlation between overdensity and neutral column
density at each redshift shown in Figure~\ref{fig: colspa}.  The scatter
increases at lower redshifts mainly because (as we will show below)
shock heated gas makes a larger contribution to the \lya forest.
Shock heating lowers the neutral fraction at a fixed density, and
the outliers in Figure~\ref{fig: colspa} almost all lie at column
densities that are lower than the mean trend.

The solid lines in Figure~\ref{fig: colspa} show a power law fitted
to the data points at each redshift.  The dotted lines show the fits 
from the other redshifts, for comparison.  The exponent remains roughly
constant with $\rho_H/\bar{\rho}_H \propto \nh^{0.7}$, 
but the overdensity associated
with a typical absorber of given $\nh$ increases significantly at low
redshifts.  The physics behind this change is the same physics
that underlies our discussion of \dndz\ evolution in \S\ref{sec: ewanal}:
the physical density associated with a given overdensity falls as
the universe expands, and the resulting decline in recombination rate
is only partly compensated by the decline in photoionization rate.
The mean trends shown in Figure~\ref{fig: colspa} can be approximately
summarized by the formula
\begin{equation}\label{eqn: colspa}
\delta_H \equiv {\rho \over \bar{\rho}_H} 
\sim 20\; \left[ {\nh\over 10^{14}\cdunits}\right] ^{0.7}\; 10^{-0.4 z}.
\end{equation}
The redshift dependence in equation~(\ref{eqn: colspa}) depends on the
UV background history --- in this case taken from \cite{haa96} ---
but it is similar in all four of our cosmological models.  The overall 
scaling factor depends on the mean flux decrement normalization, which in 
this case is based on Rauch et al.'s (1997b) determination at $z=3$.
Bryan et al.\ (1998), for example, find a significantly higher overdensity
at specified column density at $z=3$ than equation~(\ref{eqn: colspa}) implies
because their mean flux decrement is substantially lower, 
$\bar{D}=0.214$ instead of $\bar{D}=0.32$ (the Rauch et al.\ 1997b value,
which we adopt here) or the earlier estimate
$\bar{D}=0.36$ by \cite{prs93}.

Figure~\ref{fig: colspa}, and its summary in equation~(\ref{eqn: colspa}),
is the key to understanding the physical relation between the low-$z$
\lya forest and the high-$z$ \lya forest.  The dynamical state of
an absorber --- expanding or collapsing, unshocked or shocked ---
depends mainly on its overdensity.  A specified column density range
picks out absorbers of progressively higher overdensity and progressively
more advanced dynamical states as the universe expands.
A $z=0$ \lya absorber is physically analogous {\it not} to a $z=3$ \lya absorber
of the same column density but to a $z=3$ absorber with column density
$10^{0.4 \times 3 / 0.7} \approx 50$ times higher.  In a less
quantitative form, this transformation could be seen already in 
Figures~\ref{fig: color} and~\ref{fig: contour}, where the different
redshift panels
seem to show similar underlying structure with a steadily shifting
color map, corresponding to the changing relation between
overdensity and column density.\footnote{For a similar reason,
the relation between the $z=1$ and $z=3$ panels of Figure~\ref{fig: color}
is reminiscent of the relation between the HI and HeII opacity maps
at $z=2.33$ shown in figure~2 of Croft et al.\ (1997b).}

To put these results in a specific context, it is useful to recall
that the $W_r \sim 0.24$\AA\ equivalent width limit typical of 
the Key Project FOS spectra corresponds to $\nh \sim 10^{14}\cdunits$
for a $b$-parameter of $30\kms$.  The Key Project spectra
identify many lines weaker than 0.24\AA\ as well, but the largest 
uniform detection limit sample is for this equivalent width limit.
At $z=3$, this column density would be produced by gas near the cosmic 
mean density, but at $z=0$ it requires gas of overdensity
$\delta_H \approx 20$, which corresponds to 
$\nh \sim 10^{15.5}-10^{16}\cdunits$ at $z=3$.
Higher resolution spectra measured with GHRS can detect lines down
to $\nh \sim 10^{13}\cdunits$, corresponding to an overdensity of
a few at $z=0$.  Future observations using STIS, and
eventually COS, should be able to
quantify the population of \lya absorbers tracing these
low-density regions quite accurately.

\subsection{Evolution of Gas Phases}\label{sec: nhtemp}

\begin{figure*}
\centerline{
\epsfxsize=5.3truein
\epsfbox[65 200 550 740]{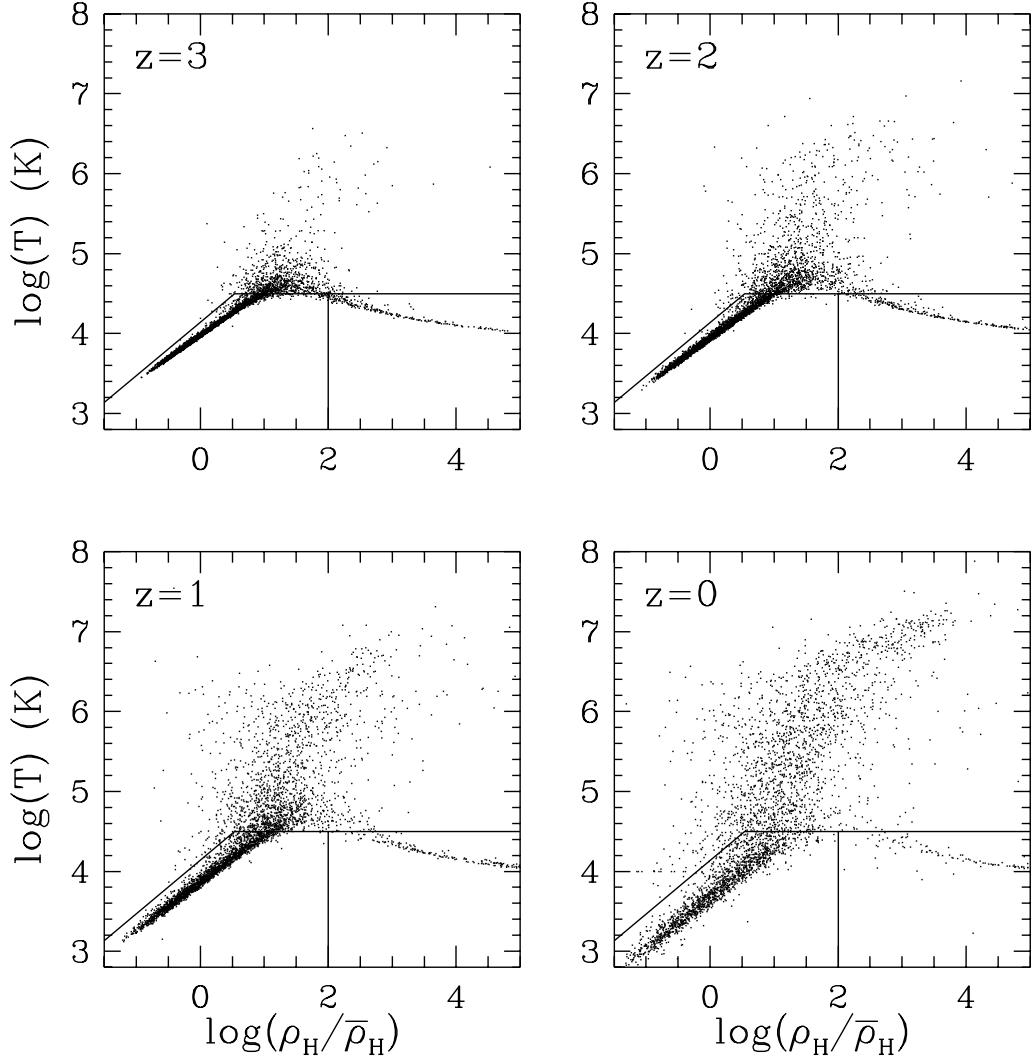}
}
\caption{              
\label{fig: nhtemp}
Temperature vs. overdensity at $z=3,2,1,$ and 0,
in the LCDM model.  The plot shows a randomly selected, $1/50^{\rm th}$
subset of the gas particles.
The gas resides in three general phases: cool,
photoionized, diffuse IGM (tail to lower left), shocked IGM, and
condensed (tail to lower right).  Lines indicate approximate demarcations
between these phases.
}
\end{figure*}

Figure~\ref{fig: nhtemp} plots the positions of SPH particles (a $1/50^{\rm th}$ subset)
from the LCDM simulation in the overdensity--temperature plane,
at $z=3,2,1,$ and 0.  At each redshift, the gas is distributed among
three broadly defined phases (\cite{kat96}):
\begin{enumerate}
\item ``Diffuse"-- a cool, low density phase associated with gas whose 
temperature is determined by the competition 
between photoionization heating and adiabatic cooling;
\item ``Shocked"-- a hot, intermediate density phase consisting of 
shock-heated gas in filaments and in galaxy, group, and cluster halos; and
\item ``Condensed"-- a cold, dense phase associated with gas in galaxies.
\end{enumerate}
Gravitational collapse moves gas from the diffuse phase to the shocked
phase, and radiative cooling moves gas from the shocked phase to the
condensed phase.

The lines in Figure~\ref{fig: nhtemp} demarcate density and
temperature regions that roughly correspond to the three phases listed above.
The lower left region corresponds to the diffuse phase, the upper
region corresponds to the shocked phase, and the lower right region
corresponds to the condensed phase.  Gas in the condensed phase is
defined to have $T<30,000$K and an overdensity greater than 100.

\begin{figure*}
\centerline{
\epsfxsize=5.3truein
\epsfbox[65 200 550 740]{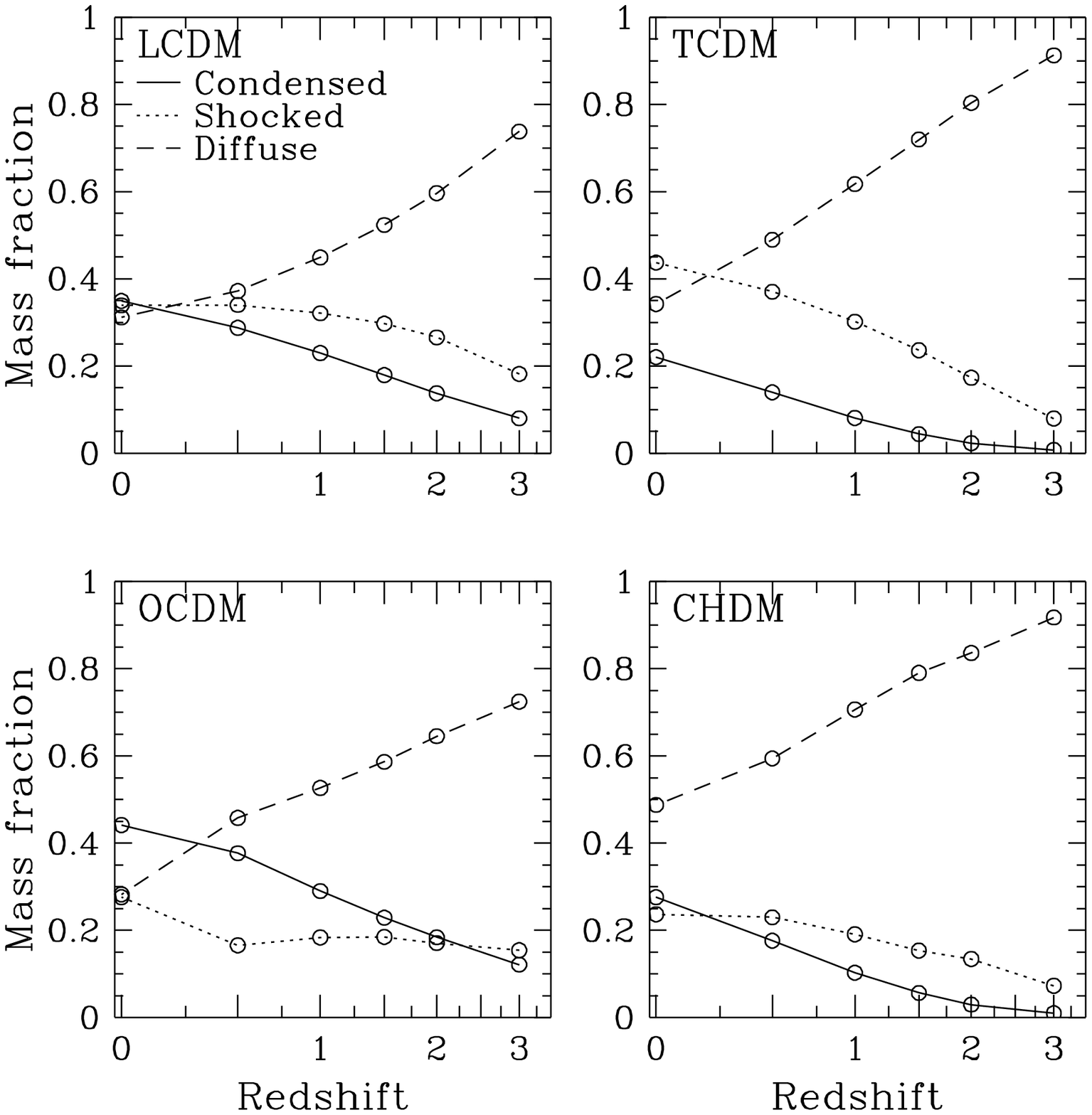}
}
\caption{              
\label{fig: ntevol}
Evolution of the baryonic mass fraction in the three
phases indicated in Fig.~\ref{fig: nhtemp}.
Above $z \sim 1$, diffuse IGM gas dominates in all the models.  
By $z=0$, mass fractions in three phases
are comparable (with some model dependence).  At all redshifts, the diffuse
component contains a significant fraction of the baryonic mass in the universe.
}
\end{figure*}

Figure~\ref{fig: ntevol} shows the evolution of the baryonic mass
fraction in each phase, in our four cosmological models.  Here we include the
stellar mass component as part of the condensed phase.  At all
redshifts, a large fraction of the baryons in the universe resides in the
diffuse phase.  By $z=0$ the fractions of baryons in the condensed and
shocked phases are comparable.  Note that the rate of evolution of
the gas is fairly model dependent, with the $\Omega=1$ models showing
more late, rapid evolution and the low-$\Omega$ models tending to
form galaxies and filamentary structure earlier.

To see how the character of the \lya forest absorbers evolves, we
examine the fraction of absorbers arising from each gas phase as a
function of redshift.  We first assign every absorber identified in our
400 random LOS a density and temperature.  This is done by determining
the velocity at which the neutral hydrogen density is maximum within $2b$ of the
absorber's central velocity (where $b$ is the $b$-parameter)
and assigning the absorber's density and
temperature to be the values at this maximum.  Using this density and
temperature and the demarcation in Figure~\ref{fig: nhtemp},
we determine the phase of gas that is predominantly
responsible for giving rise to the absorber.  

\begin{figure*}
\centerline{
\epsfxsize=5.3truein
\epsfbox[65 200 550 740]{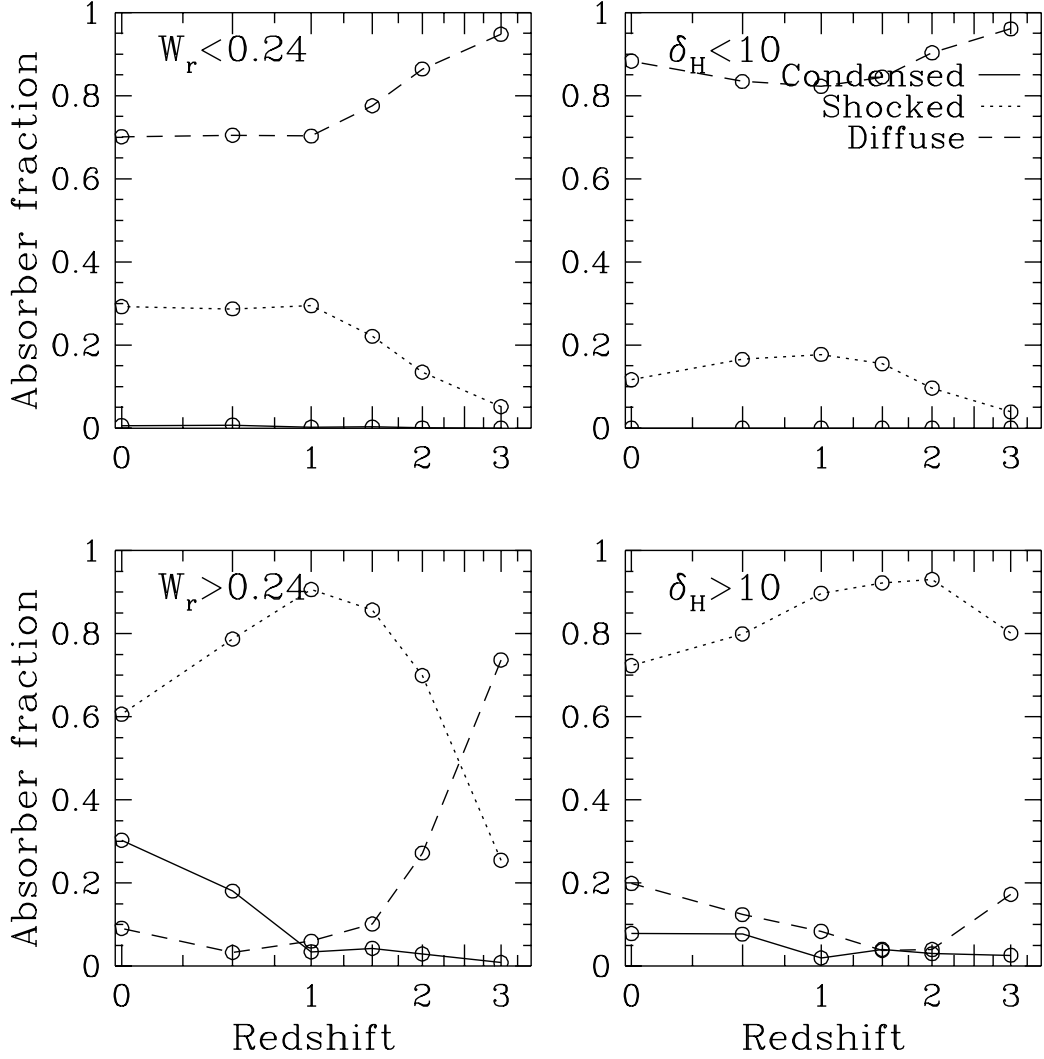}
}
\caption{              
\label{fig: abstype}
Evolution of the fraction of absorbers
associated with gas phases indicated in Fig.~\ref{fig: nhtemp},
for the LCDM model.  
{\it Left panels:} Weak absorbers with $W_r<0.24$\AA\ arise
predominantly in diffuse gas at all redshifts, while stronger absorbers
are progressively more dominated by shocked gas at lower
redshifts.  {\it Right panels:} Viewed by physical density rather than
$W_r$, absorbers show much less evolution, with low density absorbers
($\delta_H<10$) dominated by diffuse gas at all $z$ and high density
absorbers ($\delta_H>10$)
dominated by shocked gas at all $z$.  Absorption from
condensed gas in galaxies makes a small contribution at all redshifts.
}
\end{figure*}

The left panels of Figure~\ref{fig: abstype} show the 
fraction of weak ($W_r\leq 0.24$\AA) and strong
($W_r>0.24$\AA) absorbers in each of the three phases as a 
function of redshift.
The weak absorber population is dominated by diffuse gas at
all redshifts, although the contribution of shocked gas grows
with time.
The strong absorber population undergoes a marked transition
from predominantly diffuse gas at $z>2$ to predominantly 
shocked gas at $z \leq 2$.
This transition is much sharper than the change in the global
gas fractions (Figure~\ref{fig: ntevol}), and it occurs for the
reasons already discussed in \S\ref{sec: colspa}: the 0.24\AA\ 
equivalent width threshold picks out absorbers of higher overdensity
at lower redshift, and gas with $\delta_H < 10$ is typically
unshocked and gas with $\delta_H > 10$ is typically shocked
(Figure~\ref{fig: nhtemp}).
We demonstrate this point in the right hand panels of 
Figure~\ref{fig: abstype}, which show the evolution of the
phase distribution for absorbers with $\delta_H < 10$ (top)
and $\delta_H > 10$ (bottom).
With this division, there is only weak evolution in the fraction
of absorbers in different phases.  
In other words, \lya absorbers do not change character from 
high to low redshift if one selects on overdensity (which is not
directly observable), but lines of specified equivalent width 
or column density trace different physical structures at different redshifts.

Although we use the LCDM model for Figure~\ref{fig: abstype},
we find similar trends in all of our models.
Only a rather
small fraction of the strong absorbers arise from condensed gas, even
at $z=0$, although if we raised the equivalent width threshold the
fraction of condensed absorbers would increase.
Our conclusion that shocked gas begins to dominate the absorber
population at $z \la 2$ for $W_r > 0.24$\AA\ is similar to that
of Riediger et al.\ (1998, figure~6).
However, our interpretation of this result is rather different.
The cause of the transition is not that unshocked absorbers evolve
faster than shocked absorbers but that the equivalent width
threshold (or column density threshold in Riediger et al.'s analysis)
selects a different subset of the absorber population at different redshifts.

\section{Low-Redshift \lya Absorbers and Galaxies}\label{sec: gals}

\subsection{Observations}

Do \lya absorbers at low redshift arise predominantly in galaxy halos?
This question has been the focus of a number of observational studies.
\cite{lan95} imaged 46 galaxies ($0.07\la z\la 0.55$) in six fields
around Key Project spectroscopic target quasars and argued that
galaxies generally produce associated \lya absorption out to 
projected separations 
$r_p \sim 160 \hkpc$, and that at least one-third of all \lya absorbers are
physically associated with galaxies (i.e., arise from gas bound to
galaxy potentials).  \cite{che98} extended this work and found a strong
correlation of increasing equivalent width with decreasing impact
parameter to the nearest galaxy\footnote{This trend is frequently
described in the literature as an anti-correlation, but we
prefer to describe it as a correlation of increasing $W_r$ with
decreasing $r_p$ to emphasize the expected signature of increased
gas density close to galaxies.}, strengthening the case for an association 
between galaxies and \lya absorbers out to $z\sim 0.8$.  Recently \cite{tri98}
showed that this correlation extends much further than found by
\cite{che98}, to impact parameters of $\sim 500 \hkpc$.  However,
numerous observations reveal \lya absorption occurring with no nearby
galaxy (e.g., \cite{mor93}; \cite{vgo96}; \cite{bow98}), and it has been
suggested that only the stronger \lya absorbers are associated with
galaxies, whereas weaker ones routinely occur even in voids
(\cite{sto95}; \cite{shu96}; \cite{gro98}).  
A plausible population of low surface
brightness galaxies has a covering fraction that can account for \lya
absorber counts (\cite{che98}; \cite{lin98}), but deep observations
have failed to detect such galaxies around absorbers (\cite{rau96}).
Conversely, various observations suggest that \lya absorbers at low redshift
typically arise in filamentary and sheet-like networks tracing
large scale structure (Le Brun et al.\ 1996;
\cite{din97}; \cite{leb98}), similar to the structures that simulations
predict at high redshift.

While some of these findings at face value seem contradictory, the
variety of the observational procedures, selection effects, and sensitivity
make it difficult to directly compare results.  In this section, we
examine the relationship between \lya absorbers and galaxies in our
simulations, with the goal of providing interpretations for these
observations within the context of hierarchical structure
formation models.

\subsection{Identifying Galaxies}

To identify clumps of gas and stars in our simulations that are likely
to correspond to galaxies, we use the galaxy identification algorithm
SKID (\cite{kat96}; {\tt http://www-hpcc.astro.washington.edu/tools/SKID}), 
which is a modified version of
DENMAX (\cite{gel94}).  SKID identifies galaxies as gravitationally bound
groups of 16 or more cold gas ($T<30,000$~K and $\rho/\bar\rho >170$) 
and star particles that are associated with a common density maximum
(corresponding to a minimum baryon mass
$M_b \sim 1.7\times 10^9 M_\odot$ in the LCDM model). 
SKID computes a number of physical properties of these groups, including mass,
center-of-mass position and velocity, maximum circular velocity,
half-mass radius, and maximum radial extent.

We also identify the halos of dark and baryonic matter associated with
the cold, dense gas in galaxies.  We again use SKID for this purpose,
but we now apply it to all the particles in our simulations,
including dark matter, stars, and hot and cold gas.  The maximum radial
extent is the distance from the center of mass to the outermost bound
particle, and it corresponds to the bound extent of the halo.

\begin{figure*}
\centerline{
\epsfxsize=5.3truein
\epsfbox[65 200 550 740]{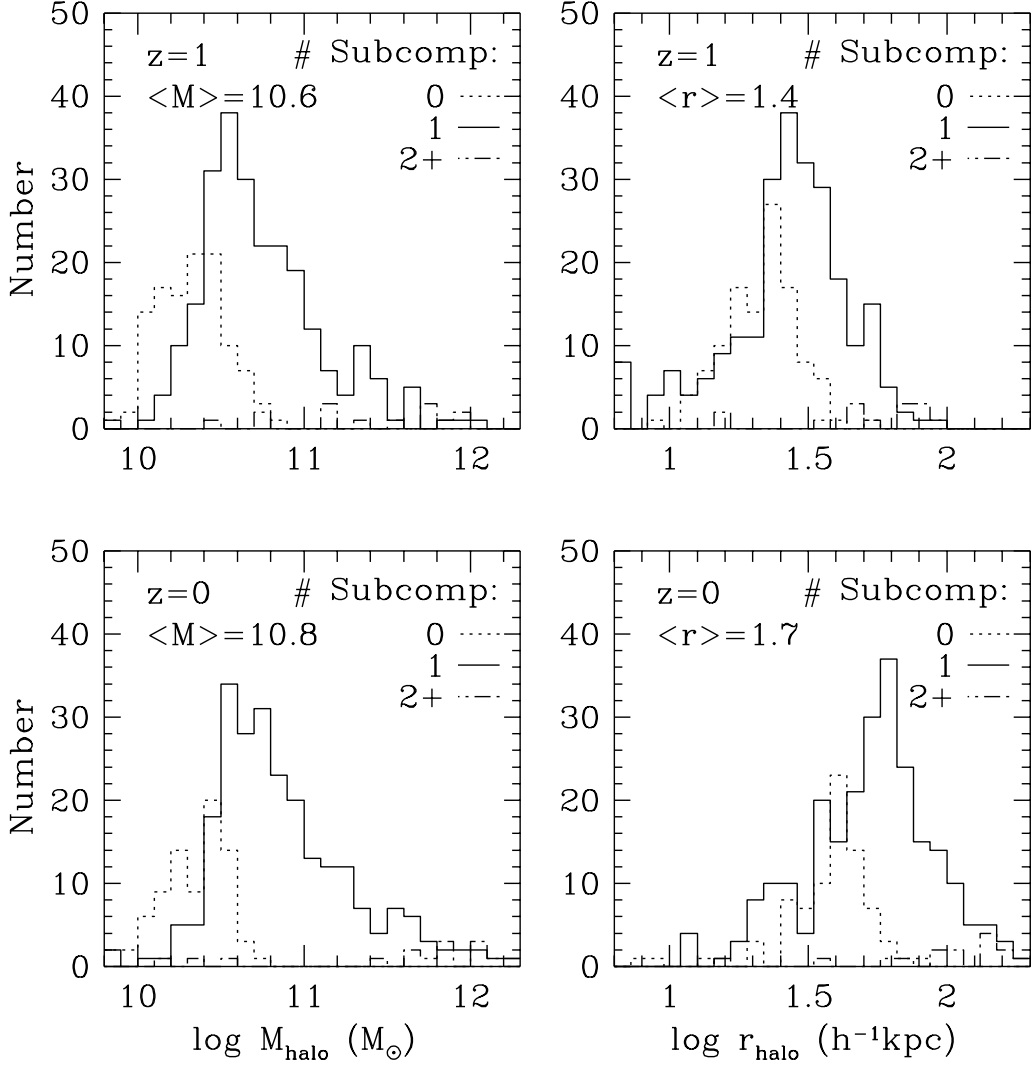}
}
\caption{              
\label{fig: subcomp}
Histograms of dark halo masses and outer
radii (bound extents), for the LCDM model at $z=1$ and $z=0$.  Left
panels show the dark halo mass, for halos with no identified baryonic
galaxy (dotted histogram), one galaxy (solid), and more than one galaxy
(dashed).  The typical halo mass (in log $M_\odot$) is shown in the
upper left.  Right panels are corresponding plots for outer halo radii
(typical value shown in $\hkpc$).
}
\end{figure*}

\begin{figure*}
\centerline{
\epsfxsize=5.3truein
\epsfbox[65 0 550 740]{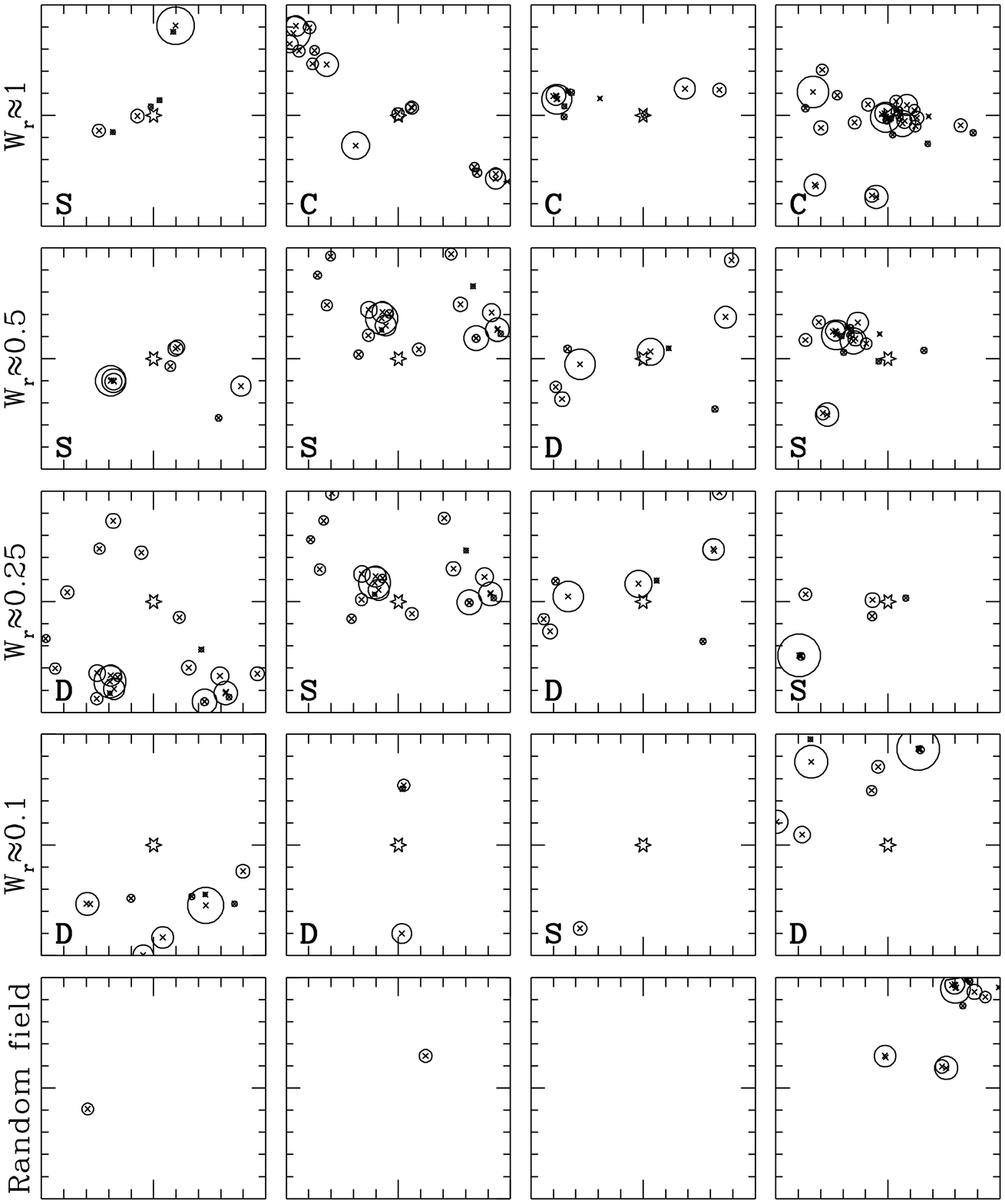}
}
\caption{              
\label{fig: fields}
Galaxy fields (2~Mpc$ \times $2~Mpc)
from the LCDM model at $z = 0$ surrounding a sample
of four random LOS from each equivalent width
range listed along the left.  The central star represents the LOS
position.  Galaxies within $200\kms$ of the absorber redshift are 
shown as crosses,
with surrounding circles indicating the maximum extent of the dark halo.
The absorber gas phase is indicated in the lower left: D=Diffuse,
S=Shocked, C=Condensed.  The bottom row shows a random sample of fields
for comparison.
}
\end{figure*}

Figure~\ref{fig: subcomp} shows distributions of the mass and bound extents
of halos with different numbers of galaxy subcomponents,
for the LCDM model at $z=1$ and $z=0$.  Some halos
have no baryonic subcomponent; this is purely a resolution effect.
A small number of halos (mostly larger, more
massive ones) have multiple galaxies.  Very rarely, a galaxy is found
by SKID that has no accompanying SKID-identified halo.  
For the majority of cases, however,
there is a one-to-one correspondence between halos and galaxies.
Figure~\ref{fig: fields} (which will be discussed in more detail in the
next section) shows the galaxies as crosses and the
(independently identified) halos as circles whose radius corresponds to
the halo's bound extent; note the good agreement between the
center-of-mass positions of galaxies and halos.

Figure~\ref{fig: subcomp} indicates that the smallest resolved
galaxies in our LCDM simulation have masses $\sim 2\times 10^{10} M_\odot$ of 
total (dark+baryonic) mass.  These galaxies are significantly smaller
and more numerous than $L^*$ galaxies.  
The halo radii are typically quite small, 
rarely exceeding $100 \hkpc$ even at $z=0$.

\subsection{A Sample of Galaxy Fields Around Absorbers}

Figure~\ref{fig: fields} shows a mosaic of 2~Mpc~$\times$~2~Mpc galaxy
fields, each centered on an absorber taken from our 400 random LOS in
the LCDM model at $z = 0$.  The central star marks the position of the absorber 
(and thus of the
artificial quasar in the background).  Crosses show galaxies identified
by SKID that are within $200\kms$ of the absorber redshift, and the
circles show the dark halos associated with these galaxies.  The circle
radius corresponds to the maximum bound extent of the halo.  
Each row shows four
absorbers that have equivalent widths close to that listed on
the left of the plot.  The absorbers in the top row are the four
strongest absorbers from the random LOS.  For comparison, the bottom
row shows four fields selected at random without regard to the presence
of an absorber; they have no central star since there is no absorber
there.  Finally, the letter in the lower left indicates the absorber's
gas phase: D=Diffuse, S=Shocked, C=Condensed.  These panels may be
compared with, for instance, Figures~3--8 of \cite{lan95} showing
imaged quasar fields.

The strongest absorbers most often arise from
condensed gas in galaxies, and, less frequently, from
shocked gas close to galaxies.  
At somewhat lower $W_r$, most of the absorbers are
associated with shocked gas in extended large scale structure around
galaxies but do not lie within the bound extent of a dark halo.
Absorbers with $W_r\la 0.3$\AA\ are typically
associated with diffuse gas and tend to lie further away from galaxy
concentrations.  Still, even at $W_r\sim 0.1$\AA, absorbers
reside in regions where there are typically more galaxies than would be
seen randomly (cf., bottom panels).  This figure suggests that while most
\lya absorbers do not arise directly from gas in galaxies, they do
tend to cluster around galaxies.  This figure and discussion are for
the LCDM model at $z=0$, but the trends are similar in all
cosmologies.

\subsection{Galaxy Slices With Lines of Sight}

\begin{figure*}
\centerline{
\epsfxsize=5.3truein
\epsfbox[65 200 550 740]{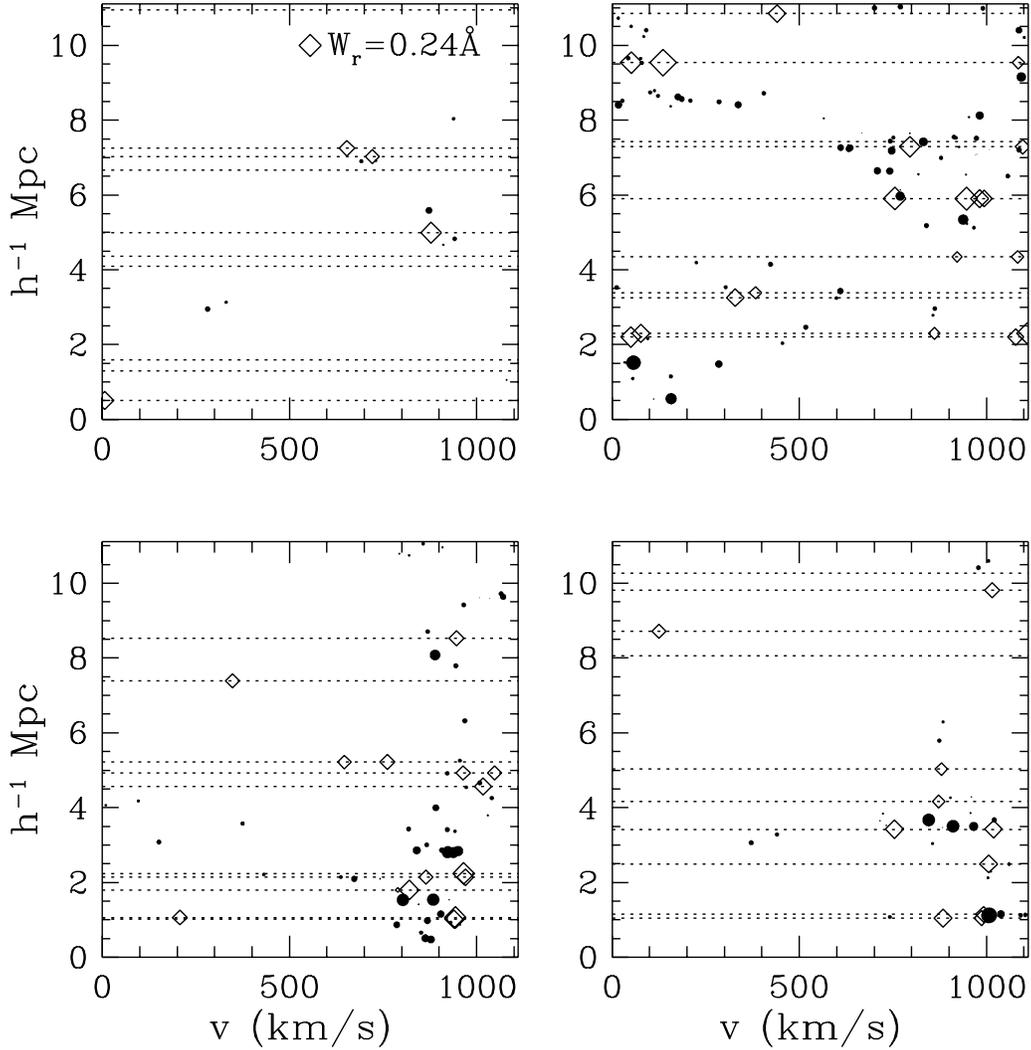}
}
\caption{              
\label{fig: slice}
Four slices (2~Mpc thick each) through
the LCDM simulation volume at $z = 0$,
with the ten random LOS nearest to the middle of the
slice shown as dotted lines.  Absorbers identified with AutoVP are
shown as diamonds, with linear size proportional to $\log(W_r)$
(see upper left panel for scale).  Galaxies are shown as filled
circles, with radius corresponding to maximum halo extent.
}
\end{figure*}

Figure~\ref{fig: slice} shows four 2~Mpc--thick slices though the LCDM
simulation volume at $z = 0$, with 10 LOS in each slice (shown as the dotted
lines).  The slices are taken along the simulation's $z$-axis (into the
page), the $y$-axis is vertical, and the $x$-axis positions of galaxies
have been shifted into redshift space using the center-of-mass velocity
of the galaxy.  The LOS are taken from the random sample of 400, and
are selected as those 10 that have $z$ positions closest to the center
of the slice.  As in Figure~\ref{fig: fields}, galaxies are shown as
circles with the radius corresponding to the dark halo's maximum bound extent.
The absorbers identified by AutoVP along each random LOS 
are shown as diamonds, with the linear size of the
diamond proportional to $\log(W_r)$.  The size representing a
$W_r=0.24$\AA\ absorber is shown in the upper left panel.  Note that
the simulation volume, and thus each panel, is periodic in every
direction.  These panels are comparable to, for example, the pie
diagrams (figure~3) in \cite{shu96}.

From Figure~\ref{fig: slice}, we see that \lya absorption occurs in a
variety of environments, with significant absorption sometimes occurring
with no nearby galaxy.  The lower left panel shows a number of
absorbers associated with the filamentary structure down the right side
of the volume.  Nearby LOS sometimes show absorbers along both LOS, and
sometimes only along one.  This is a real effect, not an artifact of
the line identification or fitting procedure.

Like Figure~\ref{fig: fields},
Figure~\ref{fig: slice} suggests that absorbers do tend to
correlate with large scale structure but that very few absorbers
arise directly from gas in galaxy halos. It also shows that some
absorbers arise in void-like regions in our simulations.

\subsection{The Absorber-Galaxy Cross-Correlation Function}\label{sec: galabs}

To quantify the visual impressions of
Figures~\ref{fig: fields} and \ref{fig: slice}, that low-redshift
\lya absorbers tend to arise from gas near but not gravitationally bound to
galaxies, we measure the cross-correlation function of galaxies and
\lya absorbers.  We define the absorber-galaxy cross-correlation function by
\begin{equation}\label{eqn: xi}
\xi(r_p,v)= {N_{\rm pairs}\over N_{\rm rand}}-1,
\end{equation}
{}where $N_{\rm pairs}$ is the number of absorber-galaxy pairs in our
simulations having projected separations between $r_p$ and
$r_p+\Delta r_p$ and line-of-sight velocity separations between $v$
and $v+\Delta v$, and $N_{\rm rand}$ is the number of pairs expected
for randomly distributed absorbers.  We find $N_{\rm pairs}$ by taking each
absorber in our 400 random LOS, finding $r_p$ and the velocity
difference to each galaxy, and binning the pairs in $100 \hkpc$ and
$100\kms$ intervals.  $N_{\rm rand}$ is determined by choosing
a random position within the simulation volume
for each absorber and performing the same procedure.

\begin{figure*}
\centerline{
\epsfxsize=5.3truein
\epsfbox[65 200 550 740]{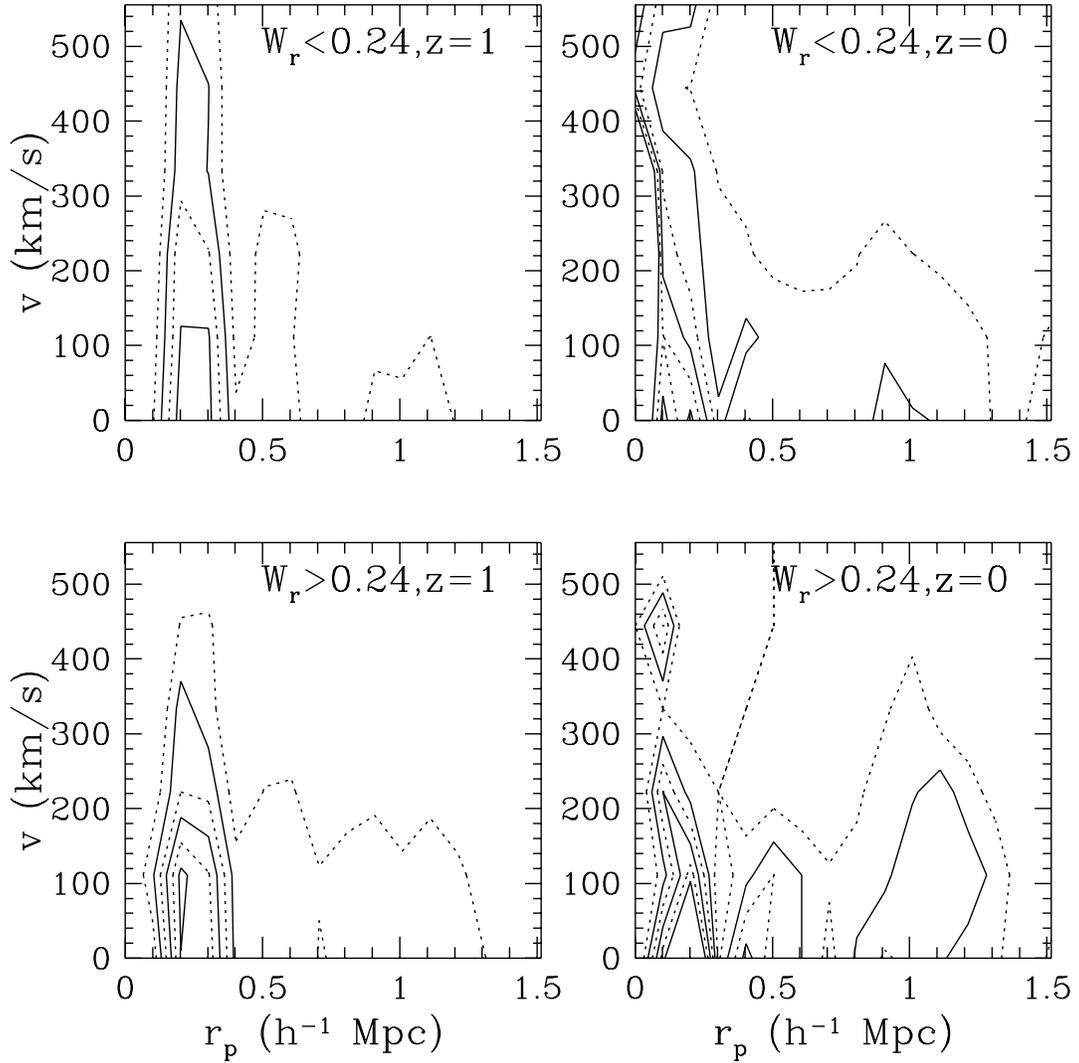}
}
\caption{              
\label{fig: galabs}
Absorber-galaxy cross-correlation function
$\xi(r_p,v)$, as defined in \S\ref{sec: galabs}, for the LCDM model:
Top and bottom panels show absorbers with $W_r<0.24$\AA\ and
$W_r>0.24$\AA, respectively.  Left and right panels are for $z=1$ and
$z=0$, respectively.  Contours show values of $\xi(r_p,v)=1,2,3,...,$
with dotted contours representing odd values and solid contours even values.
}
\end{figure*}

Figure~\ref{fig: galabs} shows $\xi(r_p,v)$ as a contour plot in $r_p$
and $v$, for absorbers with $W_r<0.24$\AA\ (upper panels) and
$W_r>0.24$\AA\ (lower panels), at $z=1$ (left panels) and $z=0$ (right
panels), for the LCDM model.  Contour levels are $\xi=1,2,3,4,5,6$,
where odd levels are dotted lines.  For instance, the outermost contour
($\xi=1$) represents the locus of $(r_p,v)$ points where 
the probability of finding a galaxy around an absorber is enhanced by 
a factor of two.  Note that the $y$-axis shows a $v$ range corresponding
to half the simulation box length, while 
$r_p$ only goes out to $1.5\hmpc$ (comoving), or 14\%
of the box length.

The value of 
$\xi(r_p,v)$ increases towards smaller velocity separations and smaller impact
parameters, as expected.  There is a decline in the number of
absorber-galaxy pairs at
small impact parameter ($r_p\la 100 \hkpc$), and to a lesser extent at
small velocity separation ($v\la 100\kms$), perhaps because the strong,
saturated absorbers near galaxies (discussed in \S\ref{sec: bindgal} below)
tend to subsume other nearby absorbers.  
In most cases, $\xi > 1$ out to projected separations of 
$r_p \sim 1\hmpc$ or more.
At small $r_p$, the correlation function exceeds unity out to
velocity separations of at least $400\kms$.  The correlation function
is highly elongated along the velocity axis because of the peculiar 
velocity of gas surrounding galaxies.
The noise in these plots is dominated by the finite number of
independent structures in our simulation volume, not by the
shot noise in the number of absorber-galaxy pairs.

The correlation function of strong absorbers (lower panels) is 
stronger than that of weak absorbers (upper panels).  For example, at $z=1$
the peak value of $\xi(r_p,v)$ is 6 for strong absorbers
but 4 for weak absorbers.  However, the difference is not dramatic,
at least for the $W_r$ division adopted here, and it appears
that the clustering of the large scale gas distribution with
galaxies gives rise to some spatial correlation for all absorbers,
regardless of their strength.  In short, \lya absorbers are correlated
significantly with galaxies out to separations of at least
$400 \hkpc$, well beyond the physical extents of the galaxies themselves.

The cross-correlation function defined by equation~(\ref{eqn: xi}) is one 
of the cleanest statistics for quantifying the relation between the 
absorber and galaxy populations.  It does not require a distinction 
between ``associated'' and ``unassociated'' pairs, and the normalization
to a random population makes it relatively insensitive
to the details of sample selection (e.g., the magnitude limit of the 
galaxy survey).  With enough data, the cross-correlation can be determined
as a function of both absorber equivalent width and galaxy luminosity.
Morris et al.\ (1993) studied a one-dimensional version of the 
cross-correlation function, combining $r_p$ and $v$ separations to
estimate an absorber-galaxy distance.
Lanzetta et al.\ (1998) have presented preliminary observational
results for $\xi(r_p,v)$; our $\xi(r_p,v)$ predictions
appear roughly consistent with these data, but perhaps somewhat weaker
at small separations.

The distribution of galaxy densities around absorbers
(\cite{gro98}) is a statistic with many of the same virtues 
as $\xi(r_p,v)$.  It responds to high-order correlations as
well as two-point correlations, and it is especially well suited
to characterizing the relations between the absorber and galaxy
populations on large scales.  Unfortunately, our current
simulations are too small to apply this statistic, at least with
the $5\hmpc$ smoothing scale adopted by Grogin \& Geller (1998).
The line-of-sight autocorrelation function of absorbers is
much more sensitive to line-blending effects than the absorber-galaxy
cross-correlation function because blending eliminates precisely
the pairs that one is attempting to count (see, e.g., Ulmer [1996]
for a measurement using the Key Project data set and for a discussion 
of blending effects).
We do not present autocorrelation results here because
even a qualitative comparison to observations requires
detailed modeling of the data properties and line identification procedure.
We suspect that statistics that treat the absorption 
spectrum as a continuous one-dimensional field
(e.g., \cite{cro97a}; \cite{rau97b}; \cite{cro98}; \cite{mir98})
will ultimately prove more powerful than line correlations for
characterizing large scale clustering in the \lya forest.

\subsection{Impact Parameter vs. Equivalent Width}\label{sec: bindgal}

The correlation of increasing equivalent width with decreasing impact
parameter is a key piece of evidence cited by \cite{che98} in support
of the hypothesis that low-redshift \lya absorbers arise in
extended gaseous envelopes of galaxies.  In order to approximately
mimic the procedure of \cite{che98}, we find the galaxy
whose outer radius (defined by the star or cold gas particle furthest from
the center of mass) has the smallest projected distance to the LOS,
for each of the 400 random LOS through the simulation.
We fit all HI features within 200$\kms$ (or, if greater, the maximum
rotational velocity) of the redshift-space velocity of
the galaxy, and sum the equivalent widths of all identified lines
\footnote{ Our results do not change significantly
if we expand our velocity interval.}. 
We identify this summed complex as a single absorber, since
current (FOS) observations are typically unable to resolve lines within
$\sim 200\kms$.  For each absorber, we find the peak neutral density
within $200\kms$ and use the gas density and temperature at this peak
to define the absorber's gas phase (cf.  \S\ref{sec: nhtemp}).  To better
sample the regions near galaxies, we generate 400 ``selected" LOS
chosen to have $10<r_p <100 \hkpc$ from a galaxy, and we apply the same
procedure.  Since the probability of a random LOS passing so close to
a galaxy is fairly small, most LOS having $r_p <100 \hkpc$ are selected
LOS.

\begin{figure*}
\centerline{
\epsfxsize=5.3truein
\epsfbox[65 50 550 740]{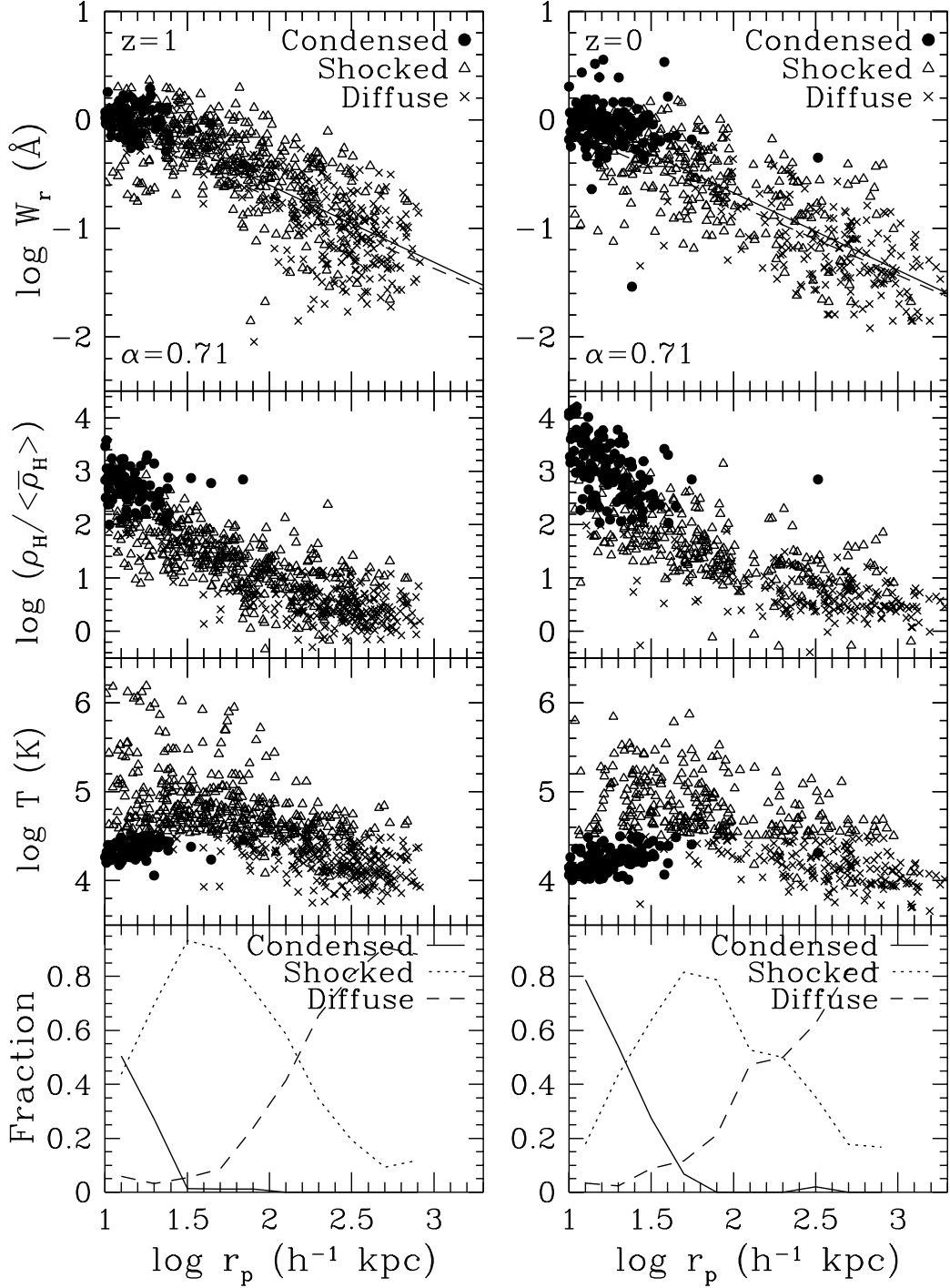}
}
\caption{              
\label{fig: gal.L64}
Equivalent width and physical properties of
absorbers vs. impact parameter, for the LCDM model.
{\it Top panels:} Scatter plot of equivalent width vs.  impact
parameter for 400 random LOS plus LOS selected to have $r_p< 100
\hkpc$, at $z=1$ and $z=0$.  Solid line is best fit to all data points;
value of $\alpha$ is shown in lower left.  Dashed line is best fit to
only diffuse IGM absorbers.  Filled circles represent condensed
absorbers, open triangles are shocked absorbers, and crosses are
diffuse absorbers.  {\it Second panels:} Overdensity at the peak of the
neutral hydrogen density within $200 \kms$ of the galaxy redshift, vs.
impact parameter.  {\it Third panels:} Temperature at the peak of the
neutral hydrogen density within $200 \kms$ of the galaxy redshift, vs.
impact parameter.  {\it Bottom panels:}
Fraction of absorbers in different phases.
}
\end{figure*}

The top panels of Figure~\ref{fig: gal.L64} show a scatter plot of
$W_r$ vs. $r_p$ for LOS drawn from the LCDM model at $z=1$ and $z=0$.
Filled circles indicate absorbers arising from condensed gas, open triangles
indicate shocked gas, and crosses indicate diffuse gas
(as defined by the demarcation of phases in Figure~\ref{fig: nhtemp}).

Following \cite{che98}, we parameterize the relation between rest
equivalent width $W_r$ and impact parameter $r_p$ by a power law,
\begin{equation}
W_r = W_{r,0} r_p^{-\alpha},
\end{equation}
{}and determine $\alpha$ from linear regression using all LOS, random
and selected.  The fit is shown as the solid line in each panel, and 
the best-fit value
of $\alpha$ is listed in the lower left.  The value of $\alpha=0.71$
does not change from $z=1$ to $z=0$ in the LCDM model.  However, it is
worth noting that the $\Omega=1$ models, which have more late structure
formation, yield $\alpha\approx 0.5$ at $z=1$ increasing to
$\alpha\approx 0.6$ at $z=0$.  The lack of significant redshift
evolution of $\alpha$ is consistent with the results of \cite{che98}.

\cite{che98} find $\alpha=0.93\pm 0.13$ for absorbers spanning
$0.1\la z\la 0.8$.  This slope is slightly higher than that inferred
from our simulations, but within $2\sigma$.  Differences in selection
procedure, spectral resolution, and other observational and numerical
details could affect a precise comparison of the data with our models.
For instance, numerical smoothing near a galaxy and unresolved ISM
clumping inside galaxy halos both serve to lower $\alpha$ in our
simulations.  In hundreds of simulated LOS we never find an
absorber with $W_r>5$\AA, whereas \cite{che98} found two such absorbers
out of 57 (see their Table~4).  Removing just these two absorbers from
their sample lowers their $\alpha$ to 0.86, so the effect can be
significant.  We also find $W_{r,0}$ in good agreement with
\cite{che98}.  Thus, all of our models reproduce the observed trend of
$W_r$ vs. $r_p$ reasonably well.

To investigate the origin of the $W_r-r_p$ correlation,
we separately determine
$\alpha$ using only those absorbers arising from diffuse gas.
This fit is shown
as the dashed line in the upper panels of Figure~\ref{fig:  gal.L64}.
In all cases, the diffuse absorber slope and overall slope are statistically
indistinguishable.  This shows that the observed correlation between
equivalent width and impact parameter can arise from the general tendency of
the large scale gas distribution to cluster around galaxies
and does not require that the \lya absorption arise in gaseous
envelopes of individual galaxies.

We expect that the $W_r-r_p$
correlation will terminate or weaken at some large impact
parameter, because eventually the statistics will be dominated by weak
absorbers whose properties are essentially uncorrelated with the galaxy 
density.  Such a lack of
correlation on large scales has been seen observationally
(\cite{shu96}), although the statistics are poor.  Unfortunately, our
simulations are too small to determine the scale where this
correlation ceases; we can only say it should extend out to $r_p \ga 500
\hkpc$ (comoving).  This prediction is in agreement with the observations
of \cite{tri98}.

In the middle two panels of Figure~\ref{fig: gal.L64} we examine
the physical state of the gas giving rise to \lya absorption near
galaxies.  The second panels show the peak gas overdensity along the
LOS within $200 \kms$ of the nearest galaxy's redshift, as a function of impact
parameter, for the LCDM model at $z=1$ and $z=0$.  The symbol types are the 
same as in the top panel.  There is a steady trend towards higher density at
lower impact parameters, down to the smallest $r_p$.  At $r_p\sim
10 \hkpc$, the overdensities can approach $10^4$, which is still lower
than the $10^6$ overdensities at the centers of the galaxies in our
simulations.  This plot clarifies the physical origin of the $W_r-r_p$
correlation: absorbers near galaxies arise in denser gas and therefore
have higher equivalent widths.  There is no clear demarcation indicating a
separation between gas in galaxies and gas in the IGM.

The third panels of Figure~\ref{fig: gal.L64} show the gas
temperatures at the peak overdensity.  There is a slight trend of
increasing temperature at smaller $r_p$ because
the diffuse gas exhibits the usual correlation between
temperature and density (Figure~\ref{fig: nhtemp}) and the fraction of 
shock-heated absorbers increases close to the galaxies.
At small impact parameters, 
a significant new population of absorbers appears, with
temperatures around $10^4$~K.  These absorbers arise directly 
in the cold, dense gas of the galaxy.

The bottom panels of Figure~\ref{fig: gal.L64} show the fraction of
absorbers at each impact parameter 
arising from condensed, shocked and diffuse gas.  Absorption
from condensed gas extends to no more than $\sim 70 \hkpc$, even at
$z=0$.  Further out, absorbers arise predominantly in shocked gas until
$r_p \sim 200 \hkpc$, and at larger $r_p$ diffuse absorbers dominate
the population.  We have also attempted to assess which absorbers are
gravitationally bound to the potential well of the nearest galaxy
and found that the majority of absorbers are bound within
$r_p \sim 50\hkpc$ and unbound at larger $r_p$.  There are very few
bound absorbers with $r_p\ga 100 \hkpc$.  These
trends are consistent with Figure~\ref{fig: subcomp}, which shows that
the dark matter halos of
galaxies in our simulation have bound extents at $z=0$ that are
typically $\sim 50 \hkpc$ and rarely exceed $\sim 100 \hkpc$.  
We find similar results for other cosmological models.

\cite{lan95} found that 10 out of their 15 sample galaxies with $r_p<160 \hkpc$ 
showed detectable \lya absorption with $W_r\ga 0.3$\AA, whereas only 1 out of 9
galaxies with $r_p>160 \hkpc$ had corresponding \lya absorption above
this threshold.  Based on this result, they argued
that most galaxies have large gaseous envelopes $\sim 160\hkpc$ in
extent and that these envelopes produce a significant fraction of
low redshift \lya forest absorbers.  We compare the simulation 
results to the \cite{lan95} findings in Table~\ref{table:
detect}.  
In the simulations, roughly $\sim 50-80\%$ of galaxies with
$r_p\leq 160\hkpc$
have an absorber within $200\kms$ with $W_r \geq 0.3$\AA, strong enough
to be detected in the line sample used by \cite{lan95}.
Conversely, for $r_p > 160 \hkpc$ the fraction drops precipitously, to
$\sim 10$\%.  The coincidence statistics and $W_r-r_p$ correlations of
\cite{lan95} and \cite{che98}
are well reproduced by our simulations, but extended gaseous envelopes
bound to galaxies are not required to explain this agreement.  We
conclude that the identification of $\sim 160 \hkpc$ as a typical
gaseous envelope size by \cite{lan95} is primarily a result
of their sample's equivalent width detection limit of $W_r\ga
0.3$\AA.  Not coincidentally, the $W_r-r_p$ relation from \cite{che98}
would predict an absorber with $r_p=160 \hkpc$ to have $W_r\approx
0.3$\AA.  A similar coincidence analysis would likely yield a smaller
inferred ``envelope'' size for a higher $W_r$ limit and a larger 
inferred size for a lower $W_r$ limit.

The hypothesis that a significant fraction of \lya absorbers reside
in galaxy halos is well motivated theoretically (\cite{bah69}),
and it is supported empirically by coincidence analyses,
by measurements of positive absorber-galaxy cross-correlations, and by
the observed $W_r-r_p$ correlation.  However, our simulations appear to
provide a good account of all of this observational evidence without
invoking extended galaxy halos, and they also account naturally
for the evidence that weaker \lya absorbers remain correlated with
galaxies but occur more often in low density environments.  Some of
the absorbers arise from condensed gas in galaxies, but these are a
small fraction of the lines with $W_r \sim 0.24$\AA\ 
(cf. Figure~\ref{fig: abstype}).  Shocked gas absorbers with
$r_p \la 50\hkpc$ are frequently bound to the gravitational potential
well of the nearest galaxy, but we have not found any sharp demarcation
in physical properties between bound and unbound absorbers, and even
the transition between shocked and diffuse absorbers that occurs
at $r_p \sim 200\hkpc$ is slow and continuous.
Setting aside the condensed absorbers,
we can characterize our results as follows:
low-redshift \lya forest absorption arises in gas that traces the
large scale structure of the underlying mass distribution, and
because galaxies tend to lie in the densest parts of the surrounding IGM,
there is a tendency for stronger absorbers to reside close to galaxies
and for weaker absorbers to reside in lower density environments.
This characterization is remarkably
similar to the ``second interpretation'' of the \lya forest 
described by \cite{tri98} (section 6), which is
the interpretation that these authors appear to favor as giving
the best overall explanation of the observational data.

\section{Discussion}\label{sec: concl}

We began our inquiry with a series of questions:
Can cosmological simulations explain the sharp transition in the evolution
of \dndz\ observed at $z \sim 1.7$?
Is the \lya forest produced by the same physical structures
at high and low redshift?
Can simulations account for the apparent correlations of
low-$z$ \lya forest absorbers with galaxies and large scale structure?

All of the cosmological models that we consider (LCDM, OCDM, TCDM, CHDM)
give a fairly good match to the observed evolution of \dndz; in particular,
all of them predict a transition from rapid evolution to slow evolution
at $z \sim 1.7$.  The ingredient crucial to this success is the
\cite{haa96} photoionizing background history, and the primary cause of the
break in \dndz\ evolution is the transition from a roughly constant
UV background intensity at $z>2$ to a rapidly declining UV background
intensity at $z<2$.  The decline in the photoionization rate counters
the decline in the recombination rate caused by expansion of the 
universe, and the cancellation of the two effects leads
to slow evolution.  Gravitational growth of structure
has a subsidiary but non-negligible effect on \dndz\ evolution,
reducing \dndz\ as gas moves from lower density regions into
collapsed structures that have smaller cross-sections for absorption.
This transformation of the underlying structure has an important
effect on the evolution of the equivalent width distribution, \dndw,
which steepens towards low redshift.  Our quantitative \dndz\ results
and our qualitative conclusions about the factors that drive the
evolution of \dndz\ at low redshift are in good agreement with those of
Theuns et al.\ (1998a) and, to a lesser extent, with those of
Riediger et al.\ (1998).

The answer to the second question --- is the \lya forest produced
by similar structures at high and low redshift --- is a more complicated
``yes and no.''  At any redshift, the population of structures producing
\lya forest absorption is physically diverse, a point emphasized by
Hernquist et al.\ (1996).  The physical properties of an absorber
are strongly correlated with its overdensity, and at any given redshift
the overdensity is in turn correlated with the absorber's defining
observable property, its neutral column density, $\nh$.
However, the correlation between column density and overdensity changes
as the universe expands 
(Figure~\ref{fig: colspa}, equation~[\ref{eqn: colspa}]), so that
an absorber with a specified $\nh$ at $z=0$ is physically analogous
to an absorber with column density $20-50$ times higher at $z=2-3$.
Structure formation drives a gradual migration of gas from 
low density regions into collapsed objects (Figure~\ref{fig: ntevol}),
but taken as a whole the population of \lya forest absorbers is still
physically similar at high and low redshift.
However, the character of absorbers at a particular value of $\nh$ changes
substantially between high and low redshift 
because a fixed column density threshold selects systems of different
overdensity at different redshifts (Figure~\ref{fig: abstype}).

The shift in the mapping between column density and overdensity
is also crucial to understanding the correlation between the \lya
forest and galaxies at low redshift.  Although we do not find
any indication of a significant population of lines produced by galaxy
halos as opposed to lines arising in the more general large scale
distribution of shocked gas, we do find that (unsurprisingly) 
galaxies tend to lie in the densest regions of the surrounding IGM.  
At $z=0-1$, the lines with column densities
$\nh \ga 10^{14}\cdunits$ (typical of the large Key Project samples)
arise in gas that is overdense by a factor of 20 or more, and
these lines tend to occur close to galaxies.
At $z=3$, gas of this overdensity would produce lines with
$\nh \sim 10^{16}\cdunits$, approaching the column density
of Lyman limit systems, which are also
closely correlated with the galaxy distribution (\cite{kat96b}, figure~2).
The moderate overdensity IGM fluctuations that produce 
$\nh \sim 10^{14}\cdunits$ 
lines at $z=2-3$ produce lines with $\nh \la 10^{13}\cdunits$
at $z=0$.  Current observational evidence suggests that these
lines are more smoothly distributed than stronger lines, as the
simulations predict.

A power law distribution of column densities has no characteristic
value, but the interplay between this distribution and the 
curve of growth (the relation between $W_r$ and $\nh$) gives a
``preferred'' status to marginally saturated lines, which for
typical $b$-parameters have\footnote{More precisely, the central 
optical depth of a Voigt-profile line is 
$\tau_c=2.52(\nh/10^{14}\cdunits)(30\kms/b)$.}
$\nh \sim 10^{14}\cdunits$.
Lines near this column density dominate the overall opacity of
the \lya forest, in the sense that they contribute the most
opacity per logarithmic interval of $\nh$.  The fully
saturated, higher column density lines are less numerous and
have only slightly larger equivalent widths, while the smaller
opacity of the lower column density lines is not fully compensated
by their larger numbers (for a power law index $\beta > 1$).
Perhaps the most significant single difference between the
high-$z$ \lya forest and the low-$z$ \lya forest is that at $z > 2$ the
``dominant'' lines are produced mainly by unshocked gas in
moderate overdensity structures (typical $\delta_H < 10$) while
at $z<1$ they are produced by shocked gas in higher overdensity
regions (typical $\delta_H \sim 10-100$).

To recap our basic results, 
the simulations successfully explain the following observed
properties of the low-$z$ \lya forest:
\begin{enumerate}
\item The break in the evolution of \dndz\ at $z \sim 1.7$ (\cite{wey98};
see Figure~\ref{fig: ewcosmo}).
\item The dependence of the evolution index $\gamma$ on equivalent width
(\cite{wey98}; see Figure~\ref{fig: ewgamma}).
\item The high abundance of weak ($W_r \la 0.1\AA$) lines detected in
high-sensitivity spectra (\cite{tri98}; a simple extrapolation from
Figure~\ref{fig: ewdist} suggests that the simulations may 
predict too many weak lines).
\item The slope, amplitude, and spatial extent of the correlation between
equivalent width and galaxy impact parameter (\cite{lan95}; \cite{che98};
\cite{tri98}; see Figure~\ref{fig: gal.L64}).
\item The fractions of galaxies with impact parameters $r_p < 160\hkpc$
and $r_p > 160\hkpc$ that have absorbers of equivalent width
$W_r > 0.3$\AA\ along the line of sight (\cite{lan95}; see 
Table~\ref{table: detect}).
\item A variety of evidence suggesting
that strong \lya absorbers are tightly correlated with galaxies
but that weaker absorbers arise preferentially in lower density
environments and occasionally in galaxy voids
(\cite{mor93}; \cite{lan95}; \cite{sto95}; Le Brun et al.\ 1996;
\cite{shu96}; \cite{vgo96}; Bowen et al.\ 1998; Lanzetta et al.\ 1998; 
\cite{leb98}; \cite{gro98}; \cite{tri98}).
\end{enumerate}

The agreement between the simulations and the observational data is
not always precise, and more closely matched analyses and comparisons
to high-sensitivity data may show that some of the currently suggestive
quantitative discrepancies are significant, at least in some of 
the cosmological models.  Nonetheless, the qualitative success
indicated by this list is remarkable because the only 
adjustable parameter in our modeling is the overall intensity scaling of
the UV background.  Furthermore, we set the value of this parameter by matching
the observed \lya flux decrement at {\it high} redshift, $z=3$, leaving all of 
the {\it low}-$z$ \lya forest
results as entirely independent predictions.  The cosmological
models themselves contain free parameters ($\Omega$, $h$, etc.),
but the values of these are set by independent cosmological considerations
that have no obvious connection to the low-$z$ \lya forest.
Although our current results do not give much handle for distinguishing
among CDM models, we view the overall success of the simulations in
reproducing the observed properties of the low-$z$ \lya forest as
strong new support for the general features of the CDM scenario.

The physical model of the low-$z$ forest that the simulations provide
is difficult to summarize concisely, because the absorber population
is diverse and the properties of the absorbers vary systematically
with column density and with environment.  In addition to the 
quantitative predictions described above, which can certainly
be tested in greater detail than we have done here, this model makes 
two key predictions that can be used to test its basic validity,
and perhaps to distinguish it from competing models that attribute
a large fraction of the low-$z$ forest to pressure-supported gas 
in mini-halos (\cite{ree86}; \cite{ike86}; \cite{mo94}) or to
low mass clouds produced by thermal instability in galactic halos
(see \cite{mo96}, who discuss this a model for Lyman limit systems).

The first prediction is that the UV background intensity 
decreases at low redshift, in roughly the fashion illustrated
in Figure~\ref{fig: matchJ}.  The required decline in $J_\nu$ is,
within current uncertainties, consistent with the predictions
of \cite{haa96} and \cite{far98}, but those predictions assume
that there are no UV background sources other than quasars,
or at least that the emissivity of any additional sources
tracks the emissivity of the quasar population.
If we assumed a constant UV background intensity from $z=2$ to $z=0$,
then the simulation predictions would change radically, and most
of the successes cited above would vanish.
Convincing evidence for a high $J_\nu$ at low redshift
(i.e., similar $J_\nu$ at $z=0$ and $z=2$) would therefore be fatal
to the physical picture that we have described.
Unfortunately, direct determinations of $J_\nu$ at low
redshift are far too uncertain to test this prediction at
present (see Figure~\ref{fig: matchJ}), and dramatically
improving the precision of these measurements may prove difficult.
The most promising route is probably to tighten the proximity
effect measurement (\cite{kul93}) using more extensive data.
In the absence of such determinations, one might choose to view
the success of this scenario of the low-$z$ \lya forest as evidence
that $J_\nu$ declines at $z<2$ and that the UV background
comes largely from quasars even at $z=0-1$ (though within current
uncertainties our models leave room for a $\sim 50\%$ contribution
from star-forming galaxies).

The second basic prediction of our scenario is a large coherence scale for
low-$z$ \lya forest absorbers, which can be tested by absorption
studies in pairs of quasars with small angular separation.
In our simulations, most \lya forest absorption arises in systems
that have low physical densities
($n_H \sim 10^{-7}{\rm cm}^{-3} - 10^{-5} {\rm cm}^{-3}$ for 
$\nh \sim 10^{13}-10^{15}\cdunits$) and low neutral fractions.
These systems must have large path lengths in order to build up
detectable neutral column densities.
Models that invoke denser gas in mini-halos or pressure-confined clouds
would typically predict much smaller sizes for individual absorbers.
We reserve detailed predictions for future work,
but it is evident from Figures~\ref{fig: color} and~\ref{fig: contour}
that absorption with $\nh \sim 10^{13}-10^{15}\cdunits$ should
show substantial coherence across lines of sight separated
by as much as several hundred $\hkpc$.  
Initial results for a quasar pair covering the absorption redshift
range $0.48 < z < 0.89$ do indicate a coherence scale of hundreds
of $\hkpc$, thus providing preliminary support for the scenario
advanced here (\cite{din97}, 1998).

Throughout this paper, we have adopted the conventional description
of the \lya forest as a collection of ``lines,'' each of which is
produced by a distinct ``absorber.''  
In several of our recent papers on the high-redshift \lya forest,
we have argued that it is often more
profitable to view a \lya forest spectrum as a continuous 1-dimensional
map of the smoothly fluctuating IGM (e.g., \cite{cro97a}b, 1998; 
\cite{rau97b}; \cite{mir98}; \cite{wei98}b). 
Statistical measures that treat the transmitted flux
as a continuous 1-dimensional field
have two practical advantages: they do not interpose a complicated line-fitting
algorithm in the path between theoretical
predictions and observational data, and they can make more complete
use of the information in an observed spectrum by detecting
the collective signature of fluctuations that are too weak
to be detected individually.
At least at high redshift, the continuous IGM view also has the
advantage of theoretical simplicity, because the \lya optical depth
is closely tied to the underlying mass density (\cite{cro97}), while
the number of \lya lines is not.
This theoretical simplicity leads to analytic and numerical approximations
that capture many of the basic results of full hydrodynamic simulations
(\cite{bi97}; Hui et al.\ 1997; \cite{cro98}; \cite{gne98}; \cite{wei98b}),
and these approximations can be used to obtain analytic lower bounds
on the baryon density (\cite{wei97}, 1998b), 
to span a wide range of cosmological parameters with
fast numerical simulations (\cite{gne97}),
to derive the power spectrum of primordial density fluctuations (\cite{cro98}),
and even to invert an observed \lya spectrum into a line-of-sight
density field (\cite{nus98}).

Unfortunately, it is not clear that the approximations that work well
at high redshift will continue to be accurate at low redshift, since
the opacity of the \lya forest at low redshift is contributed mainly
by shocked gas rather than the unshocked gas that dominates the opacity
at high redshift.  Shock heating introduces scatter into the relation
between density and temperature (see Figure~\ref{fig: nhtemp}), and
it therefore loosens the correlation between the \lya optical depth
and the underlying gas density.  The practical advantages of the
continuous field statistics still hold at low redshift,
and we suspect that such statistics will be the most
powerful tools for testing cosmological models against \lya forest
data at all redshifts.  However, the validity of techniques that
have been shown to work at high redshift must be tested anew
against hydrodynamic simulations at low redshift, and in many
cases the accuracy of these techniques is likely to decline.
We will return to these issues in future work.

There are many other opportunities for comparing these cosmological
simulations to data on the low-redshift \lya forest, including studies
of metal-line absorption to constrain the enrichment and ionization
state of different phases of the IGM (as pursued at high redshift
by \cite{hae96}; \cite{hel97}; \cite{rau97a}; \cite{dav98}),
the above-mentioned studies of absorption along paired lines
of sight (see \cite{cha97}), studies of the evolution of Lyman limit
and damped \lya systems (pursued at high redshift by \cite{kat96b};
Gardner et al. 1997ab; \cite{hae98}; \cite{ma97}), further
studies of the relationship between the absorbers and galaxies
(i.e. larger volume analogs of the simulations of
\cite{khw92}; \cite{hkw95}; \cite{whk97}; \cite{khw98}),
and studies of X-ray absorption features that
can probe higher temperature gas (\cite{cen98}; \cite{hel98a}; \cite{per98}).
The rapid development of computing technology and algorithms offers
the prospect of larger volume, higher resolution simulations of
the \lya forest in the near future.
The comparison of existing and future simulations to the increasingly
detailed and sensitive 
observations of the high- and low-redshift \lya forest will fulfill the
original promise of quasar absorption line studies, 
to place firm constraints on cosmological parameters
and to provide a
comprehensive view of structure formation and evolution from the epoch
of primeval galaxy formation to the present.

\acknowledgments
%%%%%%%%%%%%%%%%%%%%%%%%%%%%%%%%%%%%%%%%%%%%%%%%%%%%%%%%%%%%%%%%%%%%%%
We thank J. Bahcall, B. Jannuzi, and R. Weymann for stimulating
discussions, and for advance notification of their latest results.  We
also thank J. Miralda-Escud\'e for helpful comments.  We are grateful
to F.  Haardt and P. Madau for providing us with their ionizing
background in electronic form and for helpful exchanges about UV
background evolution.  This work was supported in part by the
PSC, NCSA, and SDSC supercomputing centers, by NASA theory grants
NAGW-2422, NAGW-2523, NAG5-3111, NAG5-3820, NAG5-3922, 
and NAG5-7047 by NASA LTSA grant NAG5-3525, and by the NSF under 
grants AST90-18256, ASC 93-18185, and AST 98-02568.
%%%%%%%%%%%%%%%%%%%%%%%%%%%%%%%%%%%%%%%%%%%%%%%%%%%%%%%%%%%%%%%%%%%%%%

%\appendix

\clearpage
 
\begin{deluxetable}{lcccccccc}
\footnotesize
\tablecaption{Cosmological parameters for the simulations.\label{table: models}}
\tablewidth{0pt}
\tablehead{
\colhead{Model} & 
\colhead{$\Omega$} & 
\colhead{$\Omega_\Lambda$} & 
\colhead{$\Omega_\nu$} & 
\colhead{$n$} & 
\colhead{$\Omega_b$} &
\colhead{$H_0$} &
\colhead{$\sigma_8$} &
\colhead{Age\tablenotemark{a}} 
} 
\startdata
LCDM & 0.4 & 0.6 & 0.0 & 0.95 & 0.0473 & 65 & 0.80 & 14.5 \nl
TCDM & 1.0 & 0.0 & 0.0 & 0.8 & 0.08 & 50 & 0.51 & 13.0 \nl
CHDM & 1.0 & 0.0 & 0.2 & 1.0 & 0.075 & 50 & 0.70 & 13.0 \nl
OCDM & 0.5 & 0.0 & 0.0 & 1.0 & 0.0556 & 60 & 0.72 & 12.7 \nl
\enddata
\tablenotetext{a}{In Gyr.}
\end{deluxetable}

\begin{deluxetable}{lcccccc}
\footnotesize
\tablecaption{Computational and analysis parameters for the
simulations.\label{table: cosmo}}
\tablewidth{0pt}
\tablehead{
\colhead{Model} & 
\colhead{$z_{\rm start}$} &
\colhead{$m_{\rm gas}$\tablenotemark{a}} &
\colhead{$m_{\rm dark}$\tablenotemark{a}} &
\colhead{$N_{\rm steps}$} &
\colhead{$t_{\rm CPU}$\tablenotemark{b}} &
\colhead{$f_{J_\nu}$} 
} 
\startdata
LCDM & 49 & 1.06 & 7.9 & 48800 & 17615 & 1.34 \nl
TCDM & 49 & 2.33 & 26.8 & 27656 & 7221 & 1.41 \nl
CHDM & 29 & 2.18 & 21.1\tablenotemark{c} & 15080 & 6845 & 1.50 \nl
OCDM & 99 & 1.06 & 7.9 & 30000 & ---\tablenotemark{d} & 1.77 \nl
\enddata
\tablenotetext{a}{In units of $10^8M_\odot$.}
\tablenotetext{b}{In Cray T3E node-hours.}
\tablenotetext{c}{Mass of neutrino particles is $2.91\times 10^8M_\odot$.}
\tablenotetext{d}{Run completed on a different machine.}
\end{deluxetable}

\begin{deluxetable}{lccccc}
\footnotesize
\tablecaption{Values of $\gamma$ from observations
and our four cosmological simulations.\label{table: dndz}}
\tablewidth{0pt}
\tablehead{
\colhead{$z$} & 
\colhead{Data} & 
\colhead{LCDM} & 
\colhead{TCDM} &
\colhead{OCDM} &
\colhead{CHDM} 
} 
\startdata
$3-2$   & $2.78\pm 0.71$ & 2.79 & 4.01 & 2.42 & 3.72 \nl
$2-1.5$ & unmeasured     & 1.35 & 0.90 & 1.25 & 0.42 \nl
$1.5-0$ & $0.26\pm 0.22$ & 0.91 & 0.70 & 0.62 & -0.17 \nl
\enddata
\end{deluxetable}

%\begin{deluxetable}{l|ccc|ccc}
%\footnotesize
%\tablecaption{Mean values of logarithms of halo mass and halo radius,
%in various cosmologies.
%\label{table: halos}}
%\tablewidth{0pt}
%\tablehead{
%& & $<log(M_{\rm halo})>$ & ($M_\odot$) & & $<r_{\rm halo}>$ & ($\hkpc$)\\
%\colhead{$z$} & 
%\colhead{LCDM} & 
%\colhead{TCDM} &
%\colhead{CHDM} &
%\colhead{LCDM} & 
%\colhead{TCDM} &
%\colhead{CHDM}
%} 
%\startdata
%1 & 10.76 & 11.28 & 11.00 & 30.6 & 43.8 & 35.13 \nl
%0 & ???   & 11.33 & 11.01 & ???  & 85.5 & 64.73 \nl
%\enddata
%\end{deluxetable}

\begin{deluxetable}{lccccc|ccccc}
\footnotesize
\tablecaption{Fraction of galaxies with absorbers of $W_r>0.3$\AA.\label{table: detect}}
\tablewidth{0pt}
\tablehead{
& Data & & $r_p<160$ & $\hkpc$ & & Data & & $r_p>160$ & $\hkpc$ \\
\colhead{$z$} & 
\colhead{\cite{lan95}} & 
\colhead{LCDM} & 
\colhead{TCDM} &
\colhead{OCDM} &
\colhead{CHDM} &
\colhead{\cite{lan95}} & 
\colhead{LCDM} & 
\colhead{TCDM} &
\colhead{OCDM} &
\colhead{CHDM} 
} 
\startdata
1 & 67\% & 76\% & 51\% & 77\% & 56\% & 11\% & 12\% & 16\% & 22\% & 17\% \nl
0 & 67\% & 65\% & 54\% & 45\% & 77\% & 11\% &  4\% &  3\% &  4\% & 12\% \nl
\enddata
\end{deluxetable}

%\begin{deluxetable}{lc|ccc|ccc}
%\footnotesize
%\tablecaption{Fraction of absorbers bound to galaxies.
%\label{table: assoc}}
%\tablewidth{0pt}
%\tablehead{
%& Data & & $W_r>0.3$\AA\ & & & $W_r<0.3$\AA\ \\
%\colhead{$z$} & 
%\colhead{(\cite{che98})} & 
%\colhead{LCDM} & 
%\colhead{TCDM} &
%\colhead{CHDM} &
%\colhead{LCDM} & 
%\colhead{TCDM} &
%\colhead{CHDM}
%} 
%\startdata
%1 & 32-60\% & 30\% & 1\% & 5\% & 3\% & 4\% & 2\% \nl
%0 & 32-60\% & ??? & 68\% & 43\% & ??? & 13\% & 8\% \nl
%\enddata
%\end{deluxetable}

\clearpage

%%% BIBLIOGRAPHY

% And finally, we must deal with the figures.  There are three figures
% associated with this manuscript; two figures are Encapsulated
% PostScript (EPS) files.  The third figure is a grey scale figure that does
% not exist in EPS form.
%
% Authors have three options for including figure information within a 
% manuscript.  Not all the options may be acceptable by the target Journal - be
% sure to look at the appropriate submission instructions, electronic or 
% otherwise.
%
% Option 1.  Using this option, only the figure captions are included in the
% main body of the manuscript.  The figure captions must start on a new page.
% The captions are generated with the \figcaption[]{} command: the first 
% argument is optional, if you put something in there, put the name of the 
% EPS file that goes with the caption; the second argument is the figure 
% caption itself, and may include a \label command.  The \figcaption command
% generates the figure numbers.  This option is acceptable for all manuscript
% submissions.

\clearpage


\begin{thebibliography}{}

\bibitem[Bahcall et al.\ 1991]{bah91}
    Bahcall, J. N., Jannuzi, B. T., Schneider, D. P., Hartig, G. F.,
    Bohlin, R., \& Junkkarinen, V. 1991, \apj, 377, L5
\bibitem[Bahcall \etal 1993]{bah93} Bahcall, J. N., \etal 1993, \apjs, 87, 1
\bibitem[Bahcall \etal 1996]{bah96} Bahcall, J. N., \etal 1996, \apj, 457, 19
\bibitem[Bahcall \& Salpeter 1965]{bah65} Bahcall, J. N. \& Salpeter, E. E. 
    1965, \apj, 142, 1677
\bibitem[Bahcall \& Spitzer 1969]{bah69} Bahcall, J. N., \& Spitzer, L. 1969,
     \apj, 156, L64
\bibitem[Bechtold 1994]{bec94} Bechtold, J. 1994, \apjs, 91, 1
\bibitem[{Bi }1993]{bi93}
    Bi, H.G., 1993, \apj, 405, 479
\bibitem[{Bi \& Davidsen }1997]{bi97}
    Bi, H.G., \& Davidsen, A. 1997, \apj , 479, 523
\bibitem[{Bi, Ge, \& Fang }1995]{bi95}
    Bi, H., Ge, J., \& Fang, L.-Z. 1995, \apj, 452, 90
\bibitem[Blumenthal et al.\ 1984]{blu84} Blumenthal, G. R., Faber, S. M., 
	Primack, J. R., \& Rees, M. J. 1984, Nature, 311, 517
\bibitem[Bowen, Blades, \& Pettini 1996]{bowen96}
    Bowen, D. V., Blades, J. C., \& Pettini, M., 1996, \apj, 464, 141
\bibitem[Bowen, Pettini, \& Boyle 1998]{bow98} Bowen, D. V., Pettini, M. \&
    Boyle, B. J. 1998, \mnras, 297, 239
\bibitem[Bryan et al.\ 1998]{bry98}
    Bryan, G. L., Machacek, M., Anninos, P., \& Norman, M. L. 1998,
    \apj, submitted, astro-ph/9805340
\bibitem[Burles \& Tytler 1998a]{bur98a} Burles, S. \& Tytler, D. 1998a, 
    \apj, 499, 699
\bibitem[Burles \& Tytler 1998b]{bur98b} Burles, S. \& Tytler, D. 1998b, 
    \apj, submitted, astro-ph/9712109
\bibitem[Carswell \etal 1984]{car84} Carswell, R. F., Morton, D. C., 
    Smith, M. G., Stockton, A. N., Turnshek, D. A., \& Weymann, R. J. 1984,
    \apj, 278, 486
\bibitem[Cen et al. 1994]{cen94} Cen, R., Miralda-Escud\'e, J.,
    Ostriker, J.P., \& Rauch M. 1994, \apj, 427, L9
\bibitem[Cen \& Ostriker 1998]{cen98}
    Cen, R., \& Ostriker, J. P. 1998, Science, submitted, astro-ph/9806281
\bibitem[Charlton et al.\ 1997]{cha97}
    Charlton, J. C., Anninos, P., Zhang, Y., \& Norman, M. L. 1997, \apj,
	485, 26
\bibitem[CLWB]{che98} Chen, H.-W., Lanzetta, K. M., Webb, J. K., \&
    Barcons, X. 1998, \apj, 498, 77 (\cite{che98})
\bibitem[Croft \etal 1997a]{cro97a} Croft, R. A. C., Weinberg,
    D.H., Hernquist, L., \& Katz, N. 1997a, proc. 18th Texas Symposium on 
    Relativistic Astrophysics, eds. A. Olinto, J. Frieman and D. Schramm, 
    (Singapore: World Scientific), in press, astro-ph/9701166
\bibitem[Croft et al.\ 1997b]{cro97}
    Croft, R. A. C., Weinberg, D. H., Katz, N., \& Hernquist, L. 1997b,
    \apj, 488, 532
\bibitem[Croft et al.\ 1998]{cro98}
    Croft, R. A. C., Weinberg, D. H., Katz, N., \& Hernquist, L. 1998,
    \apj, 495, 44
\bibitem[Dav\'e, Dubinski, \& Hernquist 1997a]{dav97a} Dav\'e, R., Dubinski, J.
    \& Hernquist, L. 1997a, NewAst, 2, 277
\bibitem[Dav\'e et al.\ 1998]{dav98}
    Dav\'e, R., Hellsten, U., Hernquist, L., Katz, N., \& Weinberg, D. H. 1998,
	\apj, in press, astro-ph/9803257
\bibitem[Dav\'e \etal 1997b]{dav97b} Dav\'e, R., Hernquist, L., Weinberg, 
    D.H. \& Katz, N. 1997b, \apj, 477, 21
\bibitem[Devriendt \etal 1998]{dev98} Devriendt, J. E. G., Sethi, S. K.,
    Guiderdoni, B., Nath, B. B. 1998, \mnras, in press, astro-ph/9804086
\bibitem[Dinshaw \etal 1997]{din97} Dinshaw, N., Weymann, R. J., Impey, C. D.,
    Foltz, C. B., Morris, S. L., \& Ake, T. 1997, \apj, 491, 45
\bibitem[Dinshaw \etal 1998]{din98} Dinshaw, N., Foltz, C. B.,
    Impey, C. D., \& Weymann, R. J. 1998, \apj, 494, 567
\bibitem[FGS]{far98} 
    Fardal, M., Giroux, M. L., \& Shull, J. M. 1998, 
    \apj, in press, astro-ph/9802246 (\cite{far98})
\bibitem[Gardner \etal 1997a]{gar97a} Gardner, J. P., Katz, N., 
    Hernquist, L., \& Weinberg, D. H.  1997a, \apj, 484, 31
\bibitem[Gardner \etal 1997b]{gar97b} Gardner, J. P., Katz, N., 
    Weinberg, D. H., \& Hernquist, L.,  1997b, \apj, 486, 42
\bibitem[Gelb \& Bertschinger 1994]{gel94} Gelb, J. M., \& Bertschinger, E. 1994,
    \apj, 436, 467
\bibitem[Grogin \& Geller 1998]{gro98} Grogin, N. A., \& Geller, M. J. 1998, 
    \apj, submitted, astro-ph/9804326
\bibitem[Gnedin 1997]{gne97}
    Gnedin, N. Y. 1997, \mnras, submitted, astro-ph/9706286
\bibitem[Gnedin \& Hui 1998]{gne98}
    Gnedin, N. Y., \& Hui, L. 1998, \mnras, 296, 44
\bibitem[Gunn \& Peterson 1965]{gun65} Gunn, J.E. \& Peterson, B.A. 1965,
    \apj, 142, 1633
\bibitem[HM]{haa96} Haardt, F. \& Madau, P. 1996, \apj, 461, 20 (\cite{haa96})
\bibitem[Haehnelt, Steinmetz, \& Rauch 1996]{hae96} Haehnelt, M. G., 
    Steinmetz, M. \& Rauch M. 1996, \apjl, 465, L95
\bibitem[Haehnelt, Steinmetz, \& Rauch 1998]{hae98} Haehnelt, M. G., 
    Steinmetz, M. \& Rauch M. 1998, \apj, 495, 697
\bibitem[Hellsten \etal 1997]{hel97} Hellsten, U., Dav\'e, R., Hernquist, L., 
    Weinberg, D.H. \& Katz, N. 1997, \apj, 487, 482
\bibitem[Hellsten \etal 1998]{hel98} Hellsten, U., Hernquist, L., Weinberg, D.H.
    \& Katz, N. 1998, \apj, 499, 172
\bibitem[Hellsten, Gnedin, \& Miralda-Escud\'e 1998]{hel98a}
    Hellsten, U., Gnedin, N. Y., \& Miralda-Escud\'e, J. 1998a, \apjl,
	submitted, astro-ph/9804038
\bibitem[Hernquist \& Katz 1989]{her89} Hernquist, L. \& 
    Katz, N. 1989, \apjs, 70, 419
\bibitem[Hernquist, Katz \& Weinberg 1995]{hkw95} Hernquist, L., Katz, N.,
    \& Weinberg, D.H. 1995, \apj, 442, 57
\bibitem[Hernquist \etal 1996]{her96} Hernquist, L., Katz, N., Weinberg, D.H., 
    \& Miralda-Escud\'e, J. 1996, \apjl, 457, L51
\bibitem[Hu \etal 1995]{hu95} Hu, E.M., Kim, T.S., Cowie, L.L.,
    Songaila, A., \& Rauch, M. 1995 \aj, 110, 1526 
\bibitem[Hui \& Gnedin 1997]{hg97}
    Hui, L., \& Gnedin, N. 1997, \mnras, 292, 27
\bibitem[Hui, Gnedin, \& Zhang 1997]{hui97}
    Hui, L., Gnedin, N., \& Zhang, Y. 1997, \apj, 486, 599
\bibitem[Ikeuchi 1986]{ike86}
    Ikeuchi, S. 1986, Ap \& SS, 118, 509
%\bibitem[Jannuzi 1998]{jan98} Jannuzi, B. 1998, in ``Structure
%    and Evolution of the IGM from QSO Absorption Line Systems", Proc. 13th IAP
%    Colloquium, eds. P. Petitjean and S. Charlot, 
%    (Paris: Editions Fronti\`eres), in press
\bibitem[Jannuzi \etal 1998]{jan98b} Jannuzi, B., Bahcall, J. N., 
    Bergeron, J., Boksenberg, A., Hartig, G., Kirhakos, S., Sargent, W. L. W.,
    Savage, B. D., Schneider, D. P., Turnshek, D. A., Weymann, R. J. \&
    Wolfe, A. M. 1998, \apjs, 118, in press, astro-ph/9805148
\bibitem[Katz \& Quinn 1995]{kat95} Katz, N. \& Quinn, T. 1995, TIPSY manual,
    http://www-hpcc.astro.washington.edu/tools/TIPSY
\bibitem[Katz, Hernquist \& Weinberg 1992]{khw92} Katz, N., Hernquist, L.,
    \& Weinberg D.H. 1992, \apj, 399, L109
\bibitem[Katz, Hernquist \& Weinberg 1998]{khw98} Katz, N., Hernquist, L.,
    \& Weinberg D.H. 1998, \apj, submitted, astro-ph/9806257
\bibitem[KWH]{kat96} Katz, N., Weinberg D.H., 
    \& Hernquist, L. 1996, \apjs, 105, 19 (KWH)
\bibitem[Katz et al.\ 1996b]{kat96b} Katz, N., Weinberg D.H., Hernquist, L.,
    \& Miralda-Escud\'e, J. 1996b, \apjl, 457, L57
\bibitem[KHCS]{kim97}  Kim, T.S., Hu, E.M., Cowie, L.L., \&
    Songaila, A. 1997 \aj, 114, 1 (\cite{kim97})
\bibitem[Kirkman \& Tytler 1998]{kir97} Kirkman, D. \& Tytler, D. 1998, \apj, 
    in press, astro-ph/9701209
\bibitem[Klypin, Nolthenius, \& Primack 1997]{kly97} Klypin, A., Nolthenius, R.
    \& Primack, J. 1997, \apj, 474, 533
\bibitem[Klypin \& Holtzman (1997)]{kly97b} Klypin, A. \& Holtzman, J. 1997,
    astro-ph/9712217
\bibitem[Korista, Baldwin, \& Ferland 1998]{kor98} Korista, K., Baldwin, J., \&
    Ferland, G. 1998, \apj, in press, astro-ph/9805338
\bibitem[Kulkarni \& Fall 1993]{kul93} Kulkarni, V. P. \& Fall, S. M. 1993,
    \apj, 413, 63
\bibitem[Kulkarni \etal 1996]{kul96} Kulkarni, V. P., Huang, K., Green, R. F.,
    Bechtold, J., Welty, D. \& York, D. G. 1996, \mnras, 279, 197
\bibitem[LBTW]{lan95} Lanzetta, K. M., Bowen, D. B., 
    Tytler, D. \& Webb, J. K. 1995, \apj, 442, 538 (\cite{lan95})
\bibitem[Lanzetta, Webb, \& Barcons 1998]{lan98}
    Lanzetta, K. M., Webb, J. K., \& Barcons, X. 1998, in Proc. of the 13th IAP
    Colloquium, Structure and Evolution of the IGM from QSO Absorption
    Line Systems, eds. P. Petitjean \& S. Charlot, (Paris: Editions
    Fronti\`eres), in press, astro-ph/9709168
\bibitem[Le Brun, Bergeron, \& Boiss\'e 1996]{leb96} Le Brun, V., 
    Bergeron, J. \& Boiss\'e, P. 1996, A\&A, 306, 691
\bibitem[Le Brun \& Bergeron 1998]{leb98} Le Brun, V. \& Bergeron, J.
    1998, A\&A, 332, 814
\bibitem[Liddle \etal 1996a]{lid96a} Liddle, A. R., Lyth, D. H., Roberts, D.,
    \& Viana, P. T. P. 1996a, \mnras, 278, 644
\bibitem[Liddle \etal 1996b]{lid96b} Liddle, A. R., Lyth, D. H., 
    Viana, P. T. P., \& White, M. 1996b, \mnras, 282, 281
\bibitem[Linder 1998]{lin98} Linder, S. M. 1998, \apj, 495, 637
\bibitem[Lu, Wolfe, \& Turnshek 1991]{lu91} 
    Lu, L., Wolfe, A. M., \& Turnshek, D. A. 1991, \apj, 367, L19
\bibitem[Lu \etal 1996]{lu96} Lu, L., Sargent, W.L.W., Barlow, T.A., 
    Churchill, C.W., \& Vogt, S.S. 1996, \apj, 472, 509
\bibitem[Lynds 1971]{lyn71} Lynds, R. 1971, \apj, 164, L73
\bibitem[Ma et al. 1997]{ma97} Ma, C.-P., Bertschinger, E., Hernquist,
L., Weinberg, D.H., \& Katz, N. 1997, \apj, 484, L1
\bibitem[Miralda-Escud\'e et al.\ 1996]{mir96} Miralda-Escud\'e, J.,
    Cen, R., Ostriker, J.P., \& Rauch, M. 1996, \apj, 471, 582
\bibitem[Miralda-Escud\'e \etal 1998]{mir98} Miralda-Escud\'e, J., 
    Rauch, M., Sargent, W.L.W., Barlow, T.A.,
    Weinberg, D. H., Hernquist, L., Katz, N., Cen, R., \& Ostriker, J. P.
    1998, in ``Structure
    and Evolution of the IGM from QSO Absorption Line Systems", Proc. 13th IAP
    Colloquium, eds. P. Petitjean and S. Charlot, 
    (Paris: Editions Fronti\`eres), in press, astro-ph/9710230
\bibitem[Mo \& Miralda-Escud\'e 1996]{mo96}
    Mo, H. J., \& Miralda-Escud\'e, J. 1996, \apj, 469, 589
\bibitem[Mo \& Morris 1994]{mo94} Mo, H. J. \& Morris, S. L. 1994, \mnras,
    269, 52
\bibitem[Morris \etal 1991]{mor91} Morris, S. L., Weymann, R. J., Savage, B. D.,
    \& Gilliland, R. L. 1991, \apj, 377, L21
\bibitem[Morris \etal 1993]{mor93} Morris, S. L., Weymann, R. J., Dressler, A.,
    McCarthy, P. J., Smith, B. A., Terrile, R. J., Giovanelli, R. \&
    Irwin, M. 1993, \apj, 419, 524
\bibitem[M\"ucket et al.\ 1996]{muc96}
    M\"ucket, J. P., Petitjean, P., Kates, R. E., \& Riediger, R. 1996,
    A\&A, 308, 17
\bibitem[Murdoch \etal 1986]{mur86} Murdoch, H. S., Hunstead, R. W., 
    Pettini, M., \& Blades, J. C. 1986, \apj, 309, 19
\bibitem[Nusser \& Haehnelt 1998]{nus98}
    Nusser, A., \& Haehnelt, M. 1998, \mnras, submitted, astro-ph/9806109
\bibitem[Peebles 1982]{peebles82}
    Peebles, P. J. E. 1982, \apj, 263, L1
\bibitem[Pei (1995)]{pei95} Pei, Y.-C. 1995, \apj, 438, 623
\bibitem[Perna \& Loeb 1998]{per98}
    Perna, R., \& Loeb, A. 1998, \apjl, submitted, astro-ph/9804076
\bibitem[Petitjean \etal 1993]{pet93} Petitjean, P., Webb, J. K., Rauch, M.,
    Carswell, R. F., \& Lanzetta, K. M. 1993, \mnras, 262, 499
\bibitem[Press, Rybicki, \& Schneider (1993)]{prs93} Press, W.H.,
    Rybicki, G.B., \& Schneider, D.P. 1993, \apj, 414, 64
\bibitem[Rauch, Haehnelt, \& Steinmetz 1997a]{rau97a}
    Rauch, M., Haehnelt, M. G., \& Steinmetz, M. 1997a, \apj, 481, 601
\bibitem[Rauch \etal 1997b]{rau97b} Rauch, M., Miralda-Escud\'e, J., 
    Sargent, W.L.W., Barlow, T.A., Weinberg D.H., Hernquist, L., Katz, N., 
    Cen, R., \& Ostriker, J.P. 1997b, \apj, 489, 7
\bibitem[Rauch, Weymann, \& Morris 1996]{rau96} Rauch, M., Weymann, R. J., \&
    Morris, S. L. 1996, \apj, 458, 518
\bibitem[Rees 1986]{ree86}
    Rees, M. J. 1986, \mnras, 218, L25
\bibitem[Riediger, Petitjean, \& M\"ucket 1998]{rie98} Riediger, R., 
    Petitjean, P., \& M\"ucket, J. P. 1998, A\&A, 329, 30
\bibitem[Sargent \etal 1980]{sar80} Sargent, W.L.W., Young, P.J., 
    Boksenberg, A., \& Tytler, D. 1980, \apjs, 42, 41
\bibitem[Scheuer 1965]{sche65}
    Scheuer, P. A. G. 1965, Nature, 207, 963
\bibitem[SSP]{shu96} Shull, J. M., Stocke, J. T., \&
    Penton, S. 1996, \aj, 111, 72 (\cite{shu96})
\bibitem[Stengler-Larrea \etal 1995]{ste95} Stengler-Larrea, E. A., 
    Boksenberg, A., Steidel, C. C., Sargent, W. L. W., Bahcall, J. N., 
    Bergeron, J., Hartig, G. F., Jannuzi, B. T., Kirhakos, S., Savage, B. D.,
    Schneider, D. P., Turnshek, D. A., Weymann, R. J. 1995, \apj, 444, 64
\bibitem[Stocke \etal 1995]{sto95} Stocke, J. T., Shull, J. M., Penton, S.,
    Donahue, M. \& Carilli, C. L. 1995, \apj, 451, 24
\bibitem[Theuns et al.\ 1998a]{the98} Theuns, T., Leonard, A.,
    \& Efstathiou, G. 1998a, \mnras, submitted, astro-ph/9803245
\bibitem[Theuns et al.\ 1998b]{the98b}       
   Theuns, T., Leonard, A., Efstathiou, G., Pearce, F. R., \& Thomas, P. A. 
   1998b, \mnras, submitted, astro-ph/9805119
\bibitem[TLS]{tri98} Tripp, T. M., Lu, L., \& Savage, 
    B. D. 1998, \apj, in press, astro-ph/9806036 (\cite{tri98})
\bibitem[Ulmer 1996]{ulmer96}
    Ulmer, A. 1996, \apj, 473, 110
\bibitem[van Gorkom \etal 1996]{vgo96} van Gorkom, J. H., Carilli, C. L., 
    Stocke, J. T., Perlman, E. S. \& Shull, J. M. 1996, \aj, 112, 1397
\bibitem[Vogel \etal 1995]{vog95} Vogel, S., Weymann, R., Rauch, M., \&
    Hamilton, T. 1995, \apj, 441, 162
\bibitem[Vogt et al. 1994]{vog94} Vogt, S. S., et al. 1994, SPIE, 2198, 326
\bibitem[Wadsley \& Bond 1997]{wad97}
    Wadsley, J. W., \& Bond, J.R. 1997,
    in  Proc. 12th Kingston Conference, Computational Astrophysics,
    eds. D. Clarke \& M. West, ASP Conference Series 123, (San Francisco: ASP),
    astro-ph/9612148
\bibitem[Weinberg, Hernquist \& Katz 1997]{whk97} Weinberg, D.H.,
    Hernquist, L., \& Katz, N. 1997, \apj, 477, 8
\bibitem[Weinberg et al.\ 1998a]{wei98}
    Weinberg, D.H., Hernquist, L., Katz, N., Croft, R.
    \& Miralda-Escud\'e, J.  1998a, in Proc. of the 13th IAP
    Colloquium, Structure and Evolution of the IGM from QSO Absorption
    Line Systems, eds. P. Petitjean \& S. Charlot, (Paris: Editions
    Fronti\`eres), astro-ph/9709303
\bibitem[Weinberg et al.\ 1998b]{wei98b}
    Weinberg, D. H., Katz, N., \& Hernquist, L. 1998b,
    in Origins, eds. J. M. Shull, C. E. Woodward, \& H. Thronson,
    (ASP Conference Series: San Francisco), astro-ph/9708213
\bibitem[Weinberg et al.\ 1997]{wei97}
  Weinberg, D.H., Miralda-Escud\'{e}, J., Hernquist, L., \& Katz, N., 1997,
    \apj, 490, 564
\bibitem[W98]{wey98} Weymann, R., \etal 1998, \apj, in press,
    astro-ph/9806123 (\cite{wey98})
\bibitem[White \etal 1996]{whi96} White, M., Viana, P. T. P., Liddle, A. R.
    \& Scott, D. 1996, \mnras, 283, 107
\bibitem[Zhang, Anninos, \& Norman 1995]{zha95} Zhang, Y., Anninos, P. \&
    Norman, M.L. 1995, \apjl, 453, L57
\bibitem[Zheng et al.\ 1997]{zhe97}
    Zheng, W., Kriss, G. A., Telfer, R. C., Grimes, J. P., \& Davidsen, A. F.
    1997, \apj, 475, 469


\end{thebibliography}
\end{document}